\documentclass[12pt,preprint]{aastex}
\begin{document}

\newcommand{\up}[1]{\ifmmode^{\rm #1}\else$^{\rm #1}$\fi}
\newcommand{\zdot}{\makebox[0pt][l]{.}}
\newcommand{\upd}{\up{d}}
\newcommand{\uph}{\up{h}}
\newcommand{\upm}{\up{m}}
\newcommand{\ups}{\up{s}}
\newcommand{\arcd}{\ifmmode^{\circ}\else$^{\circ}$\fi}
\newcommand{\arcm}{\ifmmode{'}\else$'$\fi}
\newcommand{\arcs}{\ifmmode{''}\else$''$\fi}

\title{Direct Distances to Cepheids in the Large Magellanic Cloud: Evidence
for a Universal Slope of the Period-Luminosity Relation up to Solar Abundance
}

\author{Wolfgang Gieren}
\affil{Universidad de Concepci{\'o}n, Departamento de Fisica,
Astronomy Group,
Casilla 160-C, Concepci{\'o}n, Chile}
\authoremail{wgieren@astro-udec.cl}
\author{Jesper Storm}
\affil{Astrophysikalisches Institut Potsdam, An der Sternwarte 16,
D-14482, Potsdam, Germany}
\authoremail{jstorm@aip.de}
\author{Thomas G. Barnes III}
\affil{The University of Texas at Austin, McDonald Observatory, 1 University Station,
C1402, Austin, TX 78712-0259, USA}
\authoremail{tgb@astro.as.utexas.edu}
\author{Pascal Fouqu{\'e}}
\affil{Observatoire Midi-Pyr{\'e}n{\'e}es, UMR 5572, 14 avenue Edouard Belin, 
F-31400 Toulouse, France}
\authoremail{pfouque@ast.obs-mip.fr}
\author{Grzegorz Pietrzy{\'n}ski}
\affil{Universidad de Concepci{\'o}n, Departamento de Fisica, Astronomy
Group, Casilla 160-C, Concepci{\'o}n, Chile}
\affil{Warsaw University Observatory, Al. Ujazdowskie 4,00-478, Warsaw,
Poland}
\authoremail{pietrzyn@hubble.cfm.udec.cl}
\author{Francesco Kienzle}
\affil{Observatoire de Geneve, 51 Ch. des Maillettes, 1290 Sauverny, Switzerland}
\authoremail{Francesco.Kienzle@obs.unige.ch}

\begin{abstract}
We have applied the infrared surface brightness (ISB) technique to derive distances to thirteen
Cepheid variables in the LMC which span a period range from 3 to 42 days. From the
absolute magnitudes of the variables calculated from these distances, we find that 
the LMC Cepheids define tight period-luminosity relations in the V, I, W, J and K bands
which agree exceedingly well with the corresponding Galactic PL relations derived 
from the same technique, and are significantly steeper than the LMC PL relations
in these bands observed by the OGLE-II Project in V, I and W, and by Persson et al. in J and K.
We find that the LMC Cepheid distance moduli we derive, after correcting them for
the tilt of the LMC bar, depend significantly on
the period of the stars, in the sense that the shortest-period Cepheids have distance
moduli near 18.3, whereas the longest-period Cepheids are found to lie near 18.6.
Since such a period dependence of the tilt-corrected LMC distance moduli should not exist, there
must be a systematic, period-dependent error in the ISB technique not discovered in
previous work. We identify
as the most likely culprit the p-factor which is used to convert the observed Cepheid
radial velocities into their pulsational velocities. By demanding i) a zero slope on
the distance modulus vs. period diagram, and ii) a zero mean difference between the
ISB and ZAMS-fitting distance moduli of a sample of well established Galactic cluster Cepheids,
we find that p=1.58 ($\pm$ 0.02) - 0.15 ($\pm$ 0.05) log P, with the
p-factor depending more strongly on Cepheid period (and thus luminosity) than
indicated by past theoretical calculations. When we re-calculate the distances
of the LMC Cepheids with the revised p-factor suggested by our data law we not only obtain consistent
distance moduli for all stars, but also decrease the slopes in the various LMC
PL relations (and particularly in the reddening-independent K and W bands)
to values which are consistent with the values observed by OGLE-II 
and Persson et al. From our 13 Cepheids, we determine the LMC distance modulus to be 18.56 $\pm$ 0.04 mag,
with an additional estimated systematic uncertainty of $\sim$0.1 mag. 
Using the same corrected p-factor law to re-determine the distances of
the Galactic Cepheids, the new Galactic PL relations
are also found consistent with the observed optical and near-infrared PL relations in the LMC.

Our main conclusion from the ISB analysis of the LMC Cepheid sample is that within current uncertainties,
there seems to be no significant difference between the slopes of
the PL relations in Milky Way and LMC. With literature data on more metal-poor systems, it seems
now possible to conclude that the {\it slope} of the Cepheid PL relation is independent
of metallicity in the broad range in [Fe/H] from -1.0 dex to solar abundance, within a small uncertainty.
The new evidence from the first ISB analysis of a sizable sample of LMC Cepheids suggests that
the previous, steeper Galactic PL relations obtained from this technique were caused 
by an underestimation of the
period dependence in the model-based p-factor law used in the previous work. We emphasize, however,
that our current results
must be substantiated by new theoretical models capable of explaining the steeper period
dependence of the p-factor law, and we will also need data on more LMC field Cepheids to rule
out remaining concerns about the validity of our current interpretation.
\end{abstract}


\section{Introduction}
Since the discovery of the Cepheid period-luminosity (PL) relation almost a hundred
years ago by Miss H. Leavitt, Cepheid variables have played a key role in the establishment
of the extragalactic distance scale. With modern telescopes and detectors, light
curves of individual Cepheid variables can be measured with good accuracy out
to distances of about 20 Mpc, as demonstrated in the HST Key Project on the Extragalactic
Distance Scale (Freedman et al. 2001). Fitting the observed PL relation in 
a galaxy to a fiducial relation calibrated in our own Milky Way Galaxy, or in the
LMC, a robust distance estimate of the program galaxy can be obtained. The HST Key
Project used such Cepheid-based distances to some 30 nearby galaxies to calibrate
far-reaching secondary methods of distance determination in these galaxies which
were then used to provide distance estimates to more remote galaxies, distant
enough for a determination of the Hubble constant being free of biases due to galaxy
peculiar velocities. An accurate determination
of the present-day expansion rate of the Universe is necessary, in turn,
to constrain other cosmological parameters with ever increasing accuracy. More stringent
constraints on $H_0$ from improved optical/near-IR work on Cepheids and secondary
distance indicators will be an important complement to constraints coming from WMAP
and the Cosmic Background Imager in the radio part of the electromagnetic spectrum.

The fiducial PL relation used in the Cepheid process of distance determination 
is obviously of key importance for
the final distance results for the program galaxies. Unfortunately, it has been 
notoriously difficult
to calibrate the Cepheid PL relation in the Milky Way Galaxy. Even using state-of-the-art
present-day telescopes and instrumentation,
Cepheid variables
in the Milky Way, with the exception of the very nearest ones (e.g. Benedict et al. 2002)
 are too distant for accurate determinations of their distances with
direct geometrical methods, and therefore one has
to resort to more indirect techniques. Over the past four decades, the two most important
methods to calibrate the PL relation have been the use of ZAMS-fitting distances 
to Cepheids in open clusters and associations, and the use of Baade-Wesselink-type
techniques which take advantage of the observed variations of a Cepheid in magnitude, color
and radial velocity to derive its distance, and mean radius. While the cluster method
has suffered some degree of complication after {\it Hipparcos} studies revealed that
stellar evolution effects on the location of the ZAMS in the Hertzsprung-Russell diagram of open clusters
are stronger than
anticipated (van Leeuwen, 1999), Baade-Wesselink-type techniques have greatly
improved in accuracy through the move from the optical to the near-infrared domain,
and are now arguably the most accurate tool to measure the distances to individual
Cepheids. In particular, the infrared
surface brightness technique (hereafter ISB technique), as calibrated by 
Fouqu{\'e} \& Gieren (1997), has allowed
the derivation of distances to individual Galactic Cepheids with an accuracy of an
estimated 5 percent (Gieren, Fouqu{\'e} \& G{\'o}mez 1997), and the Galactic Cepheid
PL relation has been calibrated from this technique in optical and near-infrared bands
in recent papers of Gieren, Fouqu{\'e}
\& G{\'o}mez (1998), Fouqu{\'e}, Storm \& Gieren (2003), and most recently by
Storm et al. (2004). In this work, it was found that the slope of the Galactic PL
relation appears to be significantly {\it steeper}, in all photometric bands, than the
slope of the corresponding PL relation in the Magellanic Clouds as established 
by the different microlensing surveys, and
in particular by the OGLE-II survey (Udalski et al. 1999; Udalski 2000). 
The question is then: Is this finding real, reflecting perhaps
an effect of metallicity on the slope of the PL relation, or is there some
hitherto unknown systematic error in the ISB distance results for the Galactic Cepheids
which causes the  PL slope as determined from the ISB distances to be steeper than 
the one observed in the Magellanic Clouds, which is extremely well established
from many hundreds of Cepheids in both LMC and SMC.
The investigation of this question is of the utmost importance for the use of Cepheids
as distance indicators, given that many nearby spiral galaxies, including a subsample
of the galaxy sample used by the HST Key Project team, is of near-solar or even
super-solar metallicity. Using for these galaxies the LMC OGLE-II PL relation as the
fiducial in the distance determination would lead to systematic errors in the
distance moduli of these metal-rich galaxies by several tenths of a magnitude, 
depending on the period ranges
spanned by the extragalactic Cepheid samples, if the steeper Galactic slope of the 
PL relation were indeed true.

One straightforward way to check on the validity of our Galactic ISB Cepheid distance
results is to apply the technique directly on a number of Cepheids in the LMC, and to
construct from these data a PL relation in the LMC, which can be compared to the
Galactic relation. Such a comparison should shed some light on the question of the
universality of the PL relation, and is the purpose of this study. 
Some time ago, we set out to obtain
the necessary high-quality data for the application of the ISB technique
on a number of Cepheid variables in the LMC. Optical
photometry of a sample of long-period field Cepheids was obtained by Moffett et al. (1998).
We also obtained complete datasets of optical and near-IR (JK) photometry, as well as
radial velocity curves, for a sample of short-period LMC Cepheids in the rich cluster
NGC 1866 (Storm et al. 2005). A preliminary distance determination for the Cepheid
HV 12198 in NGC 1866 was already published by Gieren et al. (2000).
For the longer-period Cepheids studied by Moffett et al. (1998),
there are excellent radial velocity curves in the literature which have been measured
with the CORAVEL instrument at La Silla (Imbert 1987). Very recently, near-infrared
light curves for these stars of excellent quality
have been published by Persson et al. (2004), completing
the datasets necessary to determine ISB distances to these objects. There are now thirteen
Cepheids in the LMC with periods from 3-42 days with excellent data for the ISB analysis.
We will derive a direct distance value for each of these Cepheids in this paper, and
we will demonstrate that the PL relation in the LMC obtained from these data is
{\it identical}, within small uncertainties, to the Galactic PL relation obtained
from exactly the same technique and precepts. Since we know the {\it true} LMC Cepheid PL
relation from the work of the OGLE group on more than 600 stars, and from the recent extensive
work of Persson et al. in near-infrared bands,
we will then investigate
the reason(s) for the discrepancy between the OGLE-II/Persson and ISB results, and identify
the p-factor used to convert the observed radial velocities of Cepheids into their pulsational
velocities as the most likely culprit for a period-dependent
systematic error in the ISB distance results. Re-calibrating the period dependence of the
p-factor with our LMC Cepheid distance data, we will show that this leads to a corrected slope
of the Galactic Cepheid PL relation which is in excellent agreement with the observed
OGLE-II PL relation in the LMC and SMC, suggesting (with data on other galaxies) that
{\it in the metallicity range from -1.0 dex to solar there is no significant variation in
the slope of the Cepheid PL relation in either of the optical or near-infrared bands}. 
We point out a number of caveats, and future work which has to be done to put this conclusion
on a firmer basis.

\section{The infrared surface brightness technique}
The central idea behind the surface brightness technique is to calibrate the relation
between the stellar surface brightness and an appropriate color index. Once such a
calibration is at hand, photometry yields the stellar angular diameter and, in
the case of a Cepheid variable, the variation of its angular diameter through its pulsation
cycle. The angular diameter curve of a Cepheid measured this way can then be combined 
with its linear displacement
curve which is obtained from an integration of the observed radial velocity curve
of the variable star. A linear regression analysis of pairs of angular diameters
and linear displacements of the stellar surface observed at the same phases yields
both the distance of the star, and its mean radius.

The surface brightness technique was originally introduced by Barnes \& Evans (1976).
Fouqu{\'e} \& Gieren (1997; hereafter FG97) provided a re-calibration of the technique providing
two major improvements: first, they used accurate interferometrically determined angular
diameters of giants and supergiants bracketting the Cepheid color range which had become
available at the time, to improve the calibration of the surface brightness-color relation, 
and second, in addition to the visual (V-R) color index
used in previous work they extended the calibration to the near-infrared (V-K) and
(J-K) colors. That such an extension of the technique to the near-infrared would 
significantly reduce the random errors in the technique was already suggested by
the previous work of Welch (1994). It was borne out by the results for the distances
and radii of a large sample of Galactic Cepheid variables measured with the V, V-K 
version of the 
ISB technique by Gieren, Fouqu{\'e} \& G{\'o}mez (1997, 1998), and more recently
by Fouqu{\'e}, Storm \& Gieren (2003), and Storm et al. (2004). In these papers,
it was demonstrated that the ISB technique seems able to produce distances and radii 
of Cepheid variables
accurate to 5 percent if the datasets used in the analyses are of very high quality.
In particular, it was already shown by Barnes et al. (1977) that the method is very
insensitive to errors in the assumed {\it reddenings} of the Cepheids. This is true for both,
the V-R, and the near-infrared V-K and J-K versions of the technique. In addition, 
in their recent
paper Storm et al. (2004) were able to demonstrate that at the present level of accuracy,
the technique is also insensitive to both metallicity and gravity variations 
in the Cepheids.

One of the principal potential systematic uncertainties of the ISB technique refers to 
the assumption that the pulsating 
Cepheid variables follow the same surface brightness-color relation as stable, 
nonpulsating giants
and supergiants. The interferometric work on Cepheid variables which has been
conducted in recent years by different groups (e.g. Nordgren et al. 2002; Kervella et al. 2004a)
has impressively shown that this is indeed the case, to a high degree of accuracy.
Nordgren et al. (2002) recalibrated
the V-K surface brightness-color relation from the interferometrically measured
mean angular diameters of a number of nearby Cepheids and found agreement with the
FG97 calibration at the 4\% level. With improved data on nine Cepheids,
and a total of 145 individual interferometric measurements for them, the visual surface
brightness vs. (V-K) relation from direct interferometric observations of Cepheids
is even closer to the FG97 relation, showing agreement at the 2 percent level
(Kervella et al. 2004b).
 Very recently, the pulsations of
the nearby Cepheid l Car, which due to its proximity and large linear diameter
has the largest angular diameter (3 marcsec) of all Galactic Cepheids, were resolved 
by the ESO VLTI with very high
accuracy, and it was demonstrated in that paper that for l Car the angular diameters
measured by interferometry agree at the 1\% level with those coming from the FG97
calibration of the ISB technique (Kervella et al. 2004c). As a result
of all these recent investigations, it seems now clear that Cepheid angular diameters,
and their variations over the pulsation cycles, can be very accurately predicted
by the surface brightness-color calibration of FG97.

The other possibly serious source of systematic uncertainty in the ISB technique 
(and in fact in any Baade-Wesselink-type technique) refers to the projection or
p-factor, which is used to convert the radial velocities into pulsational velocities
of the stellar surface. Since the p-factor scales the radius variations, it enters
directly into the derived distance of a Cepheid-if p is, say, overestimated by 5\%, the
resulting distance is 5\% too large as well. An accurate determination of the p-factor,
and its dependence on Cepheid luminosity, and hence period, is therefore crucial
for the method. 

The p-factor is not only a geometrical projection factor, but also depends on the
structure of the atmosphere, and even on the way radial velocities are measured.
First computations of the p-factor based on line profiles derived from model
Cepheid atmospheres were performed by Parsons (1972). Depending on the spectral
resolution, he found values ranging between 1.30 and 1.34 as appropriate. Hindsley
\& Bell (1986) investigated from a new set of models the value of the p-factor
appropriate if the radial velocities were measured with a Griffin-type photoelectric
radial velocity spectrometer like the CORAVEL instrument (Baranne et al. 1979). From 
their results, they argued for a slightly higher value, 1.36, for p than Parsons.
Gieren et al. (1993) noted that the models of Hindsley \& Bell actually predicted
a mild dependence of the p-factor on pulsation period, and determined 

p = 1.39 - 0.03 logP

as a reasonable approximation to the results from these models. In all subsequent work of our group, and
particularly in the work of Gieren, Fouqu{\'e} \& G{\'o}mez 1998, and Storm et al.
2004, this slightly period-dependent p-factor was used in the ISB analyses of
Cepheid variables.

In the most recent and probably so far most sophisticated approach to the problem, 
Sabbey et al. (1995) found that the p-factor may even be variable over the pulsation
cycle for one given Cepheid. This conclusion was derived from non-LTE models, whereas
LTE models gave a constant p-factor for a given Cepheid, which seems in
better agreement with the fact that for the best-observed Cepheids, the angular
diameter and linear displacement curves agree exceedingly well (e.g. Fig. 2 in
Storm et al. 2004) which should not be the case if the p-factor was indeed
significantly phase-dependent. We also believe that this excellent match between the
shapes of angular diameter and linear displacement curves for the best-observed Cepheids
is a strong indication that the ISB technique is not significantly affected by problems
with a changing distance between the atmospheric layers in which the spectral lines
and the continuum are produced. We have, however, 
found that for a sizable subsample of the Cepheids we have analyzed with the ISB technique
there is a discrepancy between the shapes of the angular diameter and linear
displacement curves in the phase range {\it near minimum radius}, which could be caused
by a significant variation of the p-factor in this phase range. As discussed
in Storm et al. (2004), our standard procedure to cope with this fact is to
exclude points near minimum radius (in the phase range 0.8-1.0)
in the ISB solutions. The work of Sabbey et al. (1995) also suggested that the
way the radial velocity is determined, and therefore the instrument and reduction
procedure employed for the radial velocity determination, can have an effect on the
final result for the p-factor appropriate for the analysis of a particular set of 
radial velocity
observations. Storm et al. (2004) were able to perform a high-precision test
on this possibility by comparing very high quality datasets for the Cepheids
U Sgr and X Cyg obtained with the CORAVEL and CfA 
spectrometers, finding that there is no measurable velocity difference between the two different high-resolution
 cross-correlation based systems, and
that at a high level of confidence the p-factor
to be used for both sets of velocity data should be the same. Therefore, the combination
of stars in a sample whose radial velocity curves have all been measured by cross-correlation with
a high-resolution spectrometer, as is the case for the LMC Cepheids studied in this paper,
 should not introduce any significant additional systematic
uncertainty in the ISB distance solutions. However, the preceding discussion
clearly shows that there is still some significant uncertainty as to the correct
value of the p-factor to be used in Baade-Wesselink type analyses of Cepheids, including 
its correct dependence on the stellar luminosity,
and thus period. We will come back to this problem in section 4 in the light of our distance
results for the LMC Cepheids analyzed in this paper.

\section{New distance solutions for LMC Cepheids}
In our ISB solutions for long-period LMC Cepheids reported in this section, we have used
exactly the same code and precepts as employed by Storm et al. (2004) in their analysis
of a sample of 34 Galactic Cepheids and 5 SMC Cepheids, and in their recent work on 
Cepheid variables in the rich LMC cluster NGC1866 (Storm et al. 2005). In particular,
we have allowed for small phase shifts to optimize the agreement between the observed
angular diameter and linear displacement curves of the Cepheids. In all our solutions, and
in a consistent way with the Galactic Cepheid analyses, we have only used data in
the 0.0-0.8 phase ranges, omitting the data near minimum radius for which there is
evidence that they could be affected by systematic problems which are not present in the
0.0-0.8 phase interval. We also remark here
that a recent determination of the distances to the 34 Galactic Cepheids analyzed by
Storm et al. (2004) with the Bayesian statistical analysis code of 
Barnes et al. (2003) (as opposed to the maximum likelihood technique used in our analysis)
has provided agreement of the two sets of distances to better than 1\% 
(Barnes et al. 2005), which is
a strong confirmation of the statistical validity of our code used for
the surface brightness distance and radius determinations.

Table 1 lists the 13 LMC Cepheids for which datasets exist to carry out
an ISB solution on them. The six short-period stars are members in the rich cluster
NGC 1866 and have already been analyzed in Storm et al. (2005).
All Cepheids lie in the period range of 3-50 days
for which the ISB method has been calibrated in our previous work. Newly determined periods
for the stars were derived by minimizing the scatter in their light curves. The reddenings
for the long-period stars were adopted from Persson et al. (2004).
The optical light curve data for the long-period variables were taken from Moffett et al. (1998)
and supplemented with data from Caldwell \& Coulson (1986) for HV12815,
 Martin \& Warren (1979) for HV879,
and Madore (1975) for HV2257. Their infrared light curves in J and K were taken 
from Persson et al. (2004),
and supplemented with data from Laney \& Stobie (1986) for HV879. Radial velocity
curves for the long-period Cepheids in our sample sample have been measured with the 
CORAVEL instrument by Imbert (1987)
for HV879, HV899, HV909, HV2257 and HV2338. For HV12815, we have adopted the radial
velocity curve obtained by Caldwell \& Coulson (1986). For HV12816, we present in Table 2 
a new and accurate series of radial velocity measurements obtained with the FEROS and
CORALIE high resolution spectrographs at La Silla (Kienzle et al. 2005, in preparation).
The resulting radial velocity curve of HV 12816 is shown in Fig. 1.
This 9.1 day Cepheid is important in our current analysis since it fills the gap
between the short-period NGC 1866 Cepheids, whose periods cluster around
3 days, and the longer-period stars of our current LMC Cepheid sample with periods
between 26 and 42 days.
All photometric datasets available for our LMC Cepheid sample are of excellent quality,
comparable to the typical quality of the datasets for the Galactic stars analyzed in Storm
et al. (2004). The radial velocity curves are of excellent quality too except the dataset
for HV12815, which is somewhat noisier than the data for the other Cepheids. 

In Table 1, we present the distance and radius results from our ISB solutions on the 13 
LMC Cepheids together with their respective uncertainties. 
Table 1 also lists the phase shifts between angular diameter
and linear displacement curves which were adopted for the various Cepheids (last column). 
While a part of these observed shifts may be due to intrinsic (unknown) causes, another part
is likely due to the fact that the radial velocity and photometric datasets for the
Cepheids were not obtained contemporaneously, which can introduce some additional small misalignments
between the angular diameter and linear displacement curves due to imperfectly known
pulsation periods (Gieren, Fouqu{\'e} \& G{\'o}mez, 1997). For all variables, the shifts
are satisfactorily small, indicating that the periods are quite well determined in all cases.
If we adopted zero phase shifts for all stars, we would slightly increase the random
uncertainties on the individual distances and radii, but the conclusions of this paper would
not be changed.
Our discussion in this paper is concerned with the distance results; the radii of the variables
will be discussed in a forthcoming paper. 

In Fig. 2 we show, as a typical example, the ISB solution on the Cepheid HV2257 in order
to demonstrate that the quality of the LMC Cepheid ISB solutions
is as good as the average quality of our Galactic Cepheid distance solutions reported 
in Storm et al. (2004).
In Table 3, we present the absolute magnitudes of the LMC Cepheids in the VIJK 
photometric bands which were
calculated from the distances of Table 1, and from the intensity-mean apparent magnitudes for
the Cepheids in these bands which were determined from the datasets listed above.
 Table 3 also displays the 
reddening-free (V-I) absolute Wesenheit magnitudes of the Cepheids (with the Wesenheit magnitudes
being defined as {$ {\rm W}={\rm V} - 2.51 \times (<{\rm V}> - <{\rm I}>))$}, and
the adopted color excesses from which the absorption corrections
were calculated, using the mean ratios of total-to-selective absorption as given in Fouqu{\'e} et
al. (2003).
In the last column, the spectroscopic
metallicities of the Cepheids are given as determined in the high-resolution studies
of Luck et al. (1998) for the Cepheids HV879, HV909, HV2257 and HV2338, and by Hill et al. (2000)
for 3 red giants in NGC1866 (we assume that the metallicity of the red giants in NGC1866
is representative for its Cepheids).

\section{The LMC Cepheid period-luminosity relation from the ISB technique}
In Figs. 3-7, we show the LMC Cepheid PL relations in the different photometric bands 
as defined from our data. Overplotted are the absolute magnitudes of 34 Galactic
Cepheids determined from the ISB technique in Storm et al. (2004), and those of an additional
two Galactic Cepheids analyzed in this paper. We also show the data for the five SMC Cepheids
which were
analyzed in Storm et al. (2004) to determine the effect of metallicity on the zero point
of the PL relation. Since the five SMC Cepheids span a very narrow range in period, however,
they are not useful for constraining the slope of the PL relation in the SMC.
Figs. 3-7
demonstrate that in {\it all} bands, the absolute magnitudes of the 13 LMC Cepheids
we have studied fit the corresponding Galactic PL relation exceedingly well, and are in
significant disagreement with the PL relations in the LMC as observed by the OGLE-II project
in VIW, and by Persson et al. (2004) in JK. The fits to the Cepheid absolute magnitudes
in both the LMC and Milky Way bear this out. The slopes of the corresponding PL relations
in the Milky Way and the LMC, as given in Table 4, are nearly identical and agree to a small fraction of their
respective uncertainties, whereas the ISB-determined PL slopes in the LMC disagree by 4-5 sigma with
the slopes observed in VIWJK by OGLE-II, and Persson et al., which is clearly significant.
It should be stressed that the slopes of the LMC PL relations 
are rather precisely determined from the ISB distances of the present 13 LMC Cepheids, 
in spite of the relatively small number of stars available for our analysis, because the period
distribution of the stars is favorable for this purpose, and the random uncertainties
on the absolute magnitudes are small, particularly for the longest-period LMC Cepheids.

In Figure 8, we show the K-band absolute magnitudes of the Galactic, LMC and SMC Cepheids together with
the absolute magnitudes of seven Galactic Cepheids whose angular diameter curves were measured at
the ESO VLTI by Kervella et al. (2004b). Although some of these absolute magnitudes based on
interferometry have rather large uncertainties, the data fit very nicely on the Galactic PL relation
from the ISB technique and demonstrate the excellent agreement of the interferometrically
determined angular diameters with those from our adopted surface brightness-color relation.

 The conclusion from the existing data is then that
the LMC Cepheid PL relation, as determined from the ISB technique, turns out to be {\it identical}
to the corresponding Galactic relation, within a reasonably small uncertainty, and is significantly
at odds with the directly observed, and extremely well established PL relations 
in the LMC from the OGLE-II project in the optical bands, and from the work of Persson
et al. in the near-infrared J and K bands.

A hint to the solution of this problem comes from a comparison of the true
distance moduli of the 13 LMC Cepheids in Table 1, which show an unexpected large 
and systematic deviation in the sense that the long-period Cepheids are found on average
more distant than the short-period ones.
Since all the stars are relatively far away from the LMC bar, corrections
of their distance moduli for the tilt of the LMC plane with respect to the line of sight
are clearly important. We calculated these corrections from the geometrical model of the LMC of
van der Marel \& Cioni (2001); the values for the individual LMC Cepheids are given
in column 5 of Table 1. While the tilt corrections do alleviate the discrepancy between
the distance moduli of the short- and long-period Cepheids, a significant slope on
the distance moduli vs. period plot remains.
This is shown in Fig. 9. The observed period
dependence of the tilt-corrected LMC Cepheid distance moduli is clearly unphysical. While
there is some scatter of the moduli to be expected for different reasons (see next section),
there should clearly not exist a significant systematic trend with period. Fig. 9
then suggests that there is a systematic problem with the ISB technique which
introduces this observed period dependence. From all the sources of systematic and random uncertainty
on the ISB distances which were discussed in detail in Gieren, Fouqu{\'e} \& G{\'o}mez (1997),
there are only two sources of systematic error which can introduce such a period-dependent
systematic effect on the distances calculated with the technique. These are i) a wrong
surface brightness-color relation, and ii) a wrong conversion of radial velocity measurements
to photospheric pulsational velocities.

In the discussion in the previous section, arguments were already given that the surface
brightness-color relation in V-K is now very accurately determined. The recent interferometric
work on Cepheids has confirmed the FG97 relation at the 2\% level. To see the effect of this small
difference in the adopted surface brightness-color relation, we re-calculated the distances
of the 13 LMC Cepheids with the Kervella et al. (2004b) relation

$F_{v}$ = -0.1336 $\pm$ 0.0008 $(V-K)_{0}$ + 3.9530 $\pm$ 0.0006

replacing the FG97 relation ($F_{v}$ = -0.131 $(V-K)_{0}$ + 3.947). The result is that the LMC PL
relation becomes very slightly {\it steeper} by this modification, increasing the 
discrepancy to the observed OGLE-II/Persson PL relations in the LMC even more, and 
increasing the period dependence in the distance moduli seen in Fig. 9. Adoption of the
Kervella et al. surface brightness-color relation does therefore not alleviate the
period dependence in Fig. 9 but rather works in the opposite direction. We therefore
conclude that a wrong calibration of the Cepheid surface brightness-color relation 
can be excluded as the cause for the observed period dependence of the LMC distance moduli,
at a very high level of confidence. We are then left with the p-factor.
We therefore investigate a
recalibration of the relation between p-factor and period sufficient to reconcile the short
and long period Cepheids distances. 
We do this by i) demanding that the period
dependence of the distance moduli in Fig. 9 disappears and ii) demanding that the mean difference
between the observed ISB distances and ZAMS-fitting distances to Galactic cluster Cepheids becomes
zero, at the same time. The use of Galactic cluster Cepheids seems
to be the most reasonable approach to fix the zero point of the p-factor law in a solid empirical way.
To this end, we first have to select a sample of Galactic cluster Cepheids
with both types of distance determination. After inspecting our Galactic Cepheid database in
Storm et al. (2004), we found 12 cluster Cepheids having ISB distances, a high probability for
cluster membership (e.g. Gieren \& Fouqu{\'e} 1993; Feast 1999; Turner \& Burke 2002),
and reasonably well-determined ZAMS-fitting distances. The ZAMS-fitting
distances to these 12 stars are given in Table 5, and were adopted from Sandage et al. (2004).
They are based on a Pleiades distance modulus of 5.61, which is in excellent agreement with
the recent results of Soderblom et al. (2005) based on HST astrometry of three Pleiades member stars
(5.63 $\pm$ 0.02) and of Percival et al. (2005) based on main-sequence fitting in the near-infrared
(5.63 $\pm$ 0.05), and also consistent with the result of Southworth et al. (2005) from 
an analysis of the eclipsing binary HD23642 in the Pleiades (5.72 $\pm$ 0.05).
Since the Sandage et al. paper does not state the uncertainties of the Cepheid ZAMS-fitting distances,
 we adopted them from
Turner \& Burke (2002). Column 5 of Table 5 gives the differences of the ISB
and ZAMS-fitting moduli, in the sense ISB-ZAMS, while column 6 gives the uncertainties of
these differences, derived from a quadratic addition of the ISB and ZAMS-fitting error bars.
The mean difference between the ISB and ZAMS moduli is -0.09 mag. This is demonstrated in Fig. 10
where we plot the distance modulus differences against the period of the Cepheids. The error bars
in this diagram are dominated by the uncertainties of the ZAMS-fitting distances.

While the period dependence of the p-factor law is basically determined by the observed trend
in Fig. 9 and its zero point by the mean deviation of the cluster Cepheid ZAMS-fitting distances
from their ISB distances, the determination of both constants in the p-factor law is not orthogonal. 
After several iterations, we determined as our best {\it revised} p-factor law from our adopted approach
the following relation: 

p = 1.58 ($\pm$ 0.02) - 0.15 ($\pm$ 0.05) log P

where the uncertainties of the coefficients were derived from the observed scatter in Figs. 9 and 10.
Re-calculating the ISB distances of the cluster Cepheids in Table 5 with this revised p-factor
law, the average difference of (ISB-ZAMS) moduli now becomes zero (column 7 of Table 5); this is
demonstrated in Fig. 10 (filled circles). It is also evident from this figure that the modification 
of the p-factor law has
not introduced any significant trend of the (ISB-ZAMS) modulus differences with period, which would
hint at a problem with the newly determined period dependence in this relation. In Table 6, 
we present the revised distance moduli for the 13 LMC Cepheids calculated with the new p-factor law.
Correcting them for the tilt of the LMC bar, we obtain the values in the last column of this Table.
Plotting the distance moduli against the pulsation period in Fig. 11 demonstrates 
that the adoption of the revised
p-factor law in the calculation of the ISB distances has effectively removed any dependence of
the LMC Cepheid tilt-corrected true distance moduli on period, confirming that the revised p-factor law
both eliminates the period dependence of the ISB-calculated LMC Cepheid distance moduli, and at the
same time
produces full consistency between the set of ISB- and ZAMS-fitting distances to the Galactic cluster Cepheids,
without introducing any significant trend of the distance differences in Fig. 10 with period.

What is the effect of this re-calibration of the p-factor law on the PL relations obtained
from the ISB technique in both, the LMC and the Milky Way? To this end, we re-calculated
the ISB distances of all Milky Way and LMC Cepheids with the new p-factor law, keeping 
everything else as in the original distance calculations which produced the PL relations
in Figs. 3-7. The revised ISB distances of the Galactic Cepheids, and the resulting absolute magnitudes
in the different photometric bands are given in Table 7, which also gives the revised radii of the stars
 which will be discussed elsewhere. In Table 7, we have added 4 Cepheids to the list of Storm et al. (2004);
 the data sources we have adopted for these additional objects are given in Table 9. 
The modified PL relations now all turn out to be shallower. While the Milky
Way and LMC Cepheid PL relations from the ISB technique remain near-identical to each other in all bands,
they are now much closer to the observed OGLE-II/Persson LMC relations. The slopes of the
PL relations in LMC and Milky Way obtained with the revised p-factor law are given in
Table 8, and in Figs. 12 and 13 we show the modified PL relations in the K and W bands, respectively,
where
any remaining effect of reddening on the absolute magnitudes is basically negligible. It
is seen that {\it the change of the p-factor law has not only reconciled the distance moduli of
the short- and long-period LMC Cepheids, but has at the same time brought about very good
agreement of the ISB PL relation (in all bands) in the LMC with the observed OGLEII/Persson
relations (at the combined 1 $\sigma$ levels). It is also seen that the Milky Way PL relation
 slopes are now in very good agreement, again at the combined 1 $\sigma$ level, with the slopes of the
PL relations in the LMC, in all bands}. As the principal result of this discussion we then find
that from the requirement that the distance moduli of the LMC Cepheids cannot
 depend in a systematic way on their periods
we obtain, as a direct consequence, PL relations for both the LMC and the Milky Way
whose slopes agree very well with those of the directly observed and extremely well established
PL relations in the LMC.

\section{Discussion}
The conclusions in the previous section were based on the observed period dependence of
the tilt-corrected distance moduli of the LMC Cepheids, as shown in Fig. 9. Beyond
the significant systematic trend of the distance moduli with period, there is some appreciable scatter of
the data in this diagram, and in the corrected diagram in Fig. 12. Since the ISB distance moduli 
are very insensitive to
reddening, and to metallicity differences among the Cepheids (Welch 1994; Gieren, Fouqu{\'e} \& G{\'o}mez 1998; 
Storm et al. 2004), slight errors in the reddenings we used, or the modest
differences between the individual metallicities of the Cepheids (see data in Table 3),
are not expected to produce any significant dispersion in Figs. 9 and 11.
One of the factors which can introduce some significant random scatter in the ISB distance moduli
is the size of the amplitudes of the respective V-K color curves of the Cepheids (Gieren, Fouqu{\'e}
\& G{\'o}mez 1997). The random spread
among the distance moduli of the short-period Cepheids in our sample, which are all members of the
same cluster and therefore all at the same distance,
is probably mainly a consequence of their relatively small color amplitudes. It is more difficult to
understand the observed spread among the distance moduli of the long-period stars in our sample.
The V-K amplitude-related random errors are expected quite small for these stars
(with the exception of HV12816, which has a low color amplitude, too), as indicated by the error bars
in Figs. 9 and 11. A depth effect, in the sense that some of these Cepheids might be significantly
closer, or more distant than the LMC plane could contribute to the observed scatter, but this
seems unlikely given the young age of these stars which clearly favors their location in, or very close to the
LMC disk. An effect which will contribute to some degree to the random scatter among the distance moduli of all
Cepheids, independent of their periods, is the values of the adopted phase shifts between
the angular and linear dispacement curves in the ISB solutions. In our previous papers, we have
shown that the change of the distance modulus of a given Cepheid for any reasonable variation
of its appropriate phase shift is quite small, if the datasets are of high quality, as is the case
for all our current stars. For none of the Cepheids in this study a maximum change of its adopted
phase shift still compatible with the data would alter its distance modulus by more than $\sim$0.06 mag.
It therefore seems difficult to understand the observed random spread among the long-period Cepheid moduli
without invoking a depth effect to some degree, or some additional source of random uncertainty 
we have not identified so far. 

A possible weakness in our current interpretation of the LMC ISB distance data is the fact that all
the short-period stars are situated in the cluster NGC1866. It is not a priori exluded that NGC1866
may lie closer to us than the LMC main body, and that therefore the smaller distance moduli
of the NGC1866 Cepheids in Fig. 9 are true and not caused by a wrong p-factor applied to these stars.
Indeed, Walker et al. (2002) found from ZAMS-fitting to the cluster CMD obtained from HST data
a true distance modulus of 18.35 $\pm$ 0.05 mag, and a reddening of 0.06 mag consistent with
the canonical reddening value for the cluster used in all previous work (see Storm et al. 2005).
This value is in conflict, however, with the distance derived from red clump stars in a field around
the cluster, which is 18.53 $\pm$ 0.07 mag (Salaris et al. 2003). More recently,
Groenewegen \& Salaris (2003) used the Cepheid population of NGC1866, including the 6 Cepheids used 
in our present paper, to demonstrate that the reddening-free Wesenheit magnitudes of
the NGC1866 Cepheids are consistent with a distance difference of 0.04 $\pm$ 0.03 mag between
the cluster and the LMC disk, in the sense that the cluster is {\it more distant}
than the LMC main body by this amount. The smallness of this difference supports the idea that 
NGC1866 is very close to, or located in the LMC main body, in agreement with the result coming from
the surrounding red clump star population. The relatively young age of NGC1866 of log t $\sim$ 8.0 yr
from the theoretical pulsational period-age relation applied to its
Cepheids (Bono et al. 2005), consistent with its evolutionary age,
 would also make us expect that the cluster is located in the
LMC main body where most of the recent star formation in the LMC has taken place.
 In their study, Groenewegen \& Salaris (2003) also found evidence 
that the reddening of NGC1866 is actually somewhat higher (E(B-V)=0.12 mag) than the canonical value, and show
that the ZAMS-fitting distance to NGC1866 can be reconciled with the distance derived from the cluster
Cepheids and surrounding field red clump stars under the assumption that the higher reddening value
is correct. Such a higher reddening of the NGC1866 stars, if true, would not affect our present
ISB distance determination for them in any significant way, and it would also not 
significantly affect the LMC PL relations of this paper in the reddening-insensitive W and K bands, 
and would therefore
not alter the principal conclusions of our paper. However, it is clear from the previous
discussion that is very desirable to determine ISB distances for a number of additional
short-period Cepheids in the LMC which belong to the general field population, and not
to just one cluster, in order to make our conclusions invulnerable to the possibility that
the cluster could be located in front or behind the LMC plane by a significant amount.

From the tilt-corrected distance moduli of our LMC Cepheid sample as given in the last column of
Table 6, we derive a true LMC barycenter distance modulus of 18.56 $\pm$ 0.04 mag. To this
random uncertainty derived just from the scatter of the individual LMC Cepheid distance moduli, we
should add a systematic uncertainty which will affect the LMC distance result via the dependence of 
the adopted zero point in the p-factor law on the adopted ZAMS-fitting moduli of the Galactic cluster 
Cepheids in Table 5. This systematic uncertainty is difficult to estimate but should not exceed
0.1 mag, given that the very recent work on the Pleiades distance cited before has considerably
reduced the uncertainty on this number, to which the adopted ZAMS-fitting moduli of the Cepheids
in Table 5 are tied. As a conservative estimate, we therefore find from this work 
that $(m-M)_{0}$ (LMC)=18.56 mag, with
a 0.04 mag random and a $\sim$ 0.1 mag systematic uncertainty.
This value is in good agreement with the "canonical"
LMC distance value preferred by the HST Key Project on the Extragalactic Distance Scale (Freedman
et al. 2001). We expect that once accurate model results on the p-factor law become available,
we will be able to reduce the current systematic uncertainty on the LMC distance from the ISB
technique by tying our zero point to the models, rather than to the ZAMS-fitting scale, as we
did in our previous surface brightness distance work.

Sandage, Tammann \& Reindl (2004) (hereafter STR04) have recently analyzed Cepheid data in the LMC and have
found marginal evidence for a break in the LMC PL relation at a period of 10 days, in the
sense that Cepheids with P$<$10 days define steeper PL relations than those with
P$>$10 days. Since the OGLE-II sample contains only very few long-period Cepheids, STR04
had to enhance the OGLE-II sample with 97 long-period Cepheid whose data were taken from
a variety of sources, in order to obtain acceptable statistics in their fits. This has made
their conclusions vulnerable, however, to all the problems one can have when
combining photometric datasets in crowded fields from different sources, where differences
up to 0.1 mag for the magnitudes of the same stars are no exception. We therefore believe
that the claim of STR04, extremely important if true, has to be checked with independent high-quality
photometry of Cepheids of {\it all} periods in the LMC, up to the largest observed ones,
obtained in a very homogeneous way. Some of the authors are currently engaged
in such a new observational program which will provide very accurate Cepheid mean magnitudes
in V and I for many hundreds of variables up to periods of at least 80 days, which will be
discovered in fields in, or close to the LMC bar. This homogeneous and statistically
significant dataset will definitively prove, or disprove the claim about a period break
in the LMC PL relation. From the observational data currently at hand, and in particular
from the extremely homogeneous OGLE-II database alone, the Cepheid data seem consistent, at a
high level of confidence, with
no break of the PL slope at 10 days, or some other period.  However, as said before, this conclusion
must be checked with homogeneous photometric data on many more long-period Cepheids.
Our current ISB distance determinations of 13 LMC Cepheids are certainly fully consistent 
with no period break in the PL relation, but of course we would not see such a subtle
effect from our current small sample.

\section{Conclusions}

The derivation of direct distances to 13 LMC Cepheids 
with the ISB technique has revealed a period-dependent and significant discrepancy between the individual
distance moduli, while at the same time the PL relations in the LMC from the ISB technique
agree exceedingly well with the corresponding Milky Way Cepheid relations found in
Storm et al. (2004). Given the existing very accurate interferometric calibration of the
Cepheid surface brightness-color relation for the V, V-K magnitude/color combination
we are using in our distance analyses,
and the resulting accurate determination of the angular diameter curves of our program
Cepheids in the LMC, we identify as the most likely culprit of the period dependence
seen in the $(m-M)_{0}$ - period diagram a systematic error in the determination of the
linear displacement curves of the Cepheids from their observed radial velocity curves.
The problem is likely due to a flawed calibration of the period dependence of the p-factor
which provides the transformation of the observed radial velocities to the pulsational
velocities of the Cepheids. We find that assuming a steeper period dependence of the p-factor
law we can reconcile, within the current uncertainties, the distance moduli of the
individual LMC Cepheids (after correcting them for the tilt of the LMC plane with respect
to the line of sight with the geometrical model of van der Marel and Cioni), and the
slopes of the LMC period-luminosity relations which, using the revised p-factor law 
we derive in this study, agree to
within the combined 1 $\sigma$ uncertainties with the PL relations observed in the LMC by the OGLE-II
Project, and by Persson el al. in the near-infrared. When we re-calculate the ISB distances
of the Milky Way Cepheids with the revised p-factor relation, we obtain excellent consistency
of the slopes of the Galactic Cepheid PL relations in all optical and near-infrared bands,
and the corresponding LMC PL relations derived from the ISB technique, which all agree 
within $\pm$ 1 $\sigma$ with the directly observed and extremely well established LMC
PL relations in VIWJK. Tying the zero point of our new p-factor law calibration to a sample
of well-established Galactic cluster Cepeids, we find from our 13 LMC Cepheid sample a
true distance modulus of the LMC barycenter of 18.56, with an estimated 0.04 mag random, and
0.1 mag systematic uncertainty,
in very good agreement
with the "canonical" LMC distance adopted by the HST Key Project team.
We believe that taking all this information
together, there is now strong empirical evidence that our conclusions regarding
the need of revision of the p-factor law are correct. Evidently, we will be able to
calibrate the period dependence of the p-factor more accurately once we can obtain
ISB distance determinations for a larger sample of LMC Cepheids. A program to obtain
the necessary data, particularly high-quality radial velocity curves of selected
Cepheids, has very recently been started. It should be noted, however, that the current
sample is already providing relatively accurate information on the period dependence
of the p-factor because of the concentration of the Cepheids towards small, and long
periods. With the extended sample, we will fill in intermediate periods, and at the same
time increase the number of Cepheids with very short and very long periods.

In spite of the evidence for a need of revision of the p-factor law presented in this
paper, there are several aspects which will need further work, to definitively prove
or disprove our conclusions. First,
there is evidently a need to reproduce the "observed" p-factor law from new and refined
models of Cepheid atmospheres. Past work, as the one of Sabbey et al. (1995), has perhaps 
concentrated too much on fixing the zero point of the law and the possible variation
of p during the pulsation cycle of a Cepheid, and less on establishing its systematic
dependence on the stellar luminosity and effective temperature, and thus on the period.
Our group (Gieren et al. 1993) was the first to recognize that the models of Hindsley
and Bell (1986) actually predicted such a systematic period dependence, and the current
results of this paper at least confirm the sign of this trend (e.g. p becomes smaller
with increasing pulsation period). Given the relatively small amount of past theoretical
work invested in this problem, it is perhaps not surprising that we find a rather
strong disagreement with the observations. We also note that this discovery would
not have been possible with the study of Galactic Cepheids alone-as in the discovery
of the PL relation a hundred years ago, we needed to study a sample of Cepheids
all lying at the same distance and still being bright enough for accurate work, 
and the LMC remains the ideal place for such a study. Yet,
we cannot be completely sure if our current interpretation of the data is correct as long as
the suggested period dependence of the p factor is not physically understood from better models.
We hope that our current results will spur new investigations in this field. Another
point which needs to be improved is the inclusion of additional short-period LMC Cepheids
in our analysis which are not all members of a given cluster, as the ones we had available
for the current analysis. While we have presented arguments which in our belief support
that these Cepheids and their host cluster are actually located very close to, or in the
LMC bar, it remains a cause of concern that NGC1866 could be located in front of the
LMC bar by a significant amount, which would alter the value of the slope of the p-factor law
we derive from the data in Fig. 9, and in consequence the slopes of the PL relations in Table 8.
The acqisition of new data for the ISB analyses of such additional short-period LMC field Cepheids
is underway and should help to clarify this question.

With these caveats in mind, our current results support the evidence that the slope of the Cepheid PL relation,
at least in the optical V,I, W bands, does not vary significantly
with metallicity between -1.0 dex, and solar metallicity. Udalski et al. (2001) demonstrated
that in the metal-poor galaxy IC1613 the PL slopes agree very well with the LMC slopes,
and the same group established in the OGLE-II project that there is no measurable difference
between the PL relation slopes in LMC and SMC (at -0.7 dex) either. Recently, Gieren et al. (2004)
and Pietrzynski et al. (2004) have shown that the observed PL relations in VIW in NGC300, and NGC6822
are also extremely well fit by the respective OGLE-II LMC slope, NGC300 and NGC6822 having mean metallicities
of their young stellar populations of about -0.3 dex (Urbaneja et al. 2005), and -0.5 dex
(Venn et al. 2001), respectively. The previous indications that the slope of the PL relation
in the solar-metallicity Milky Way might be steeper than in the more metal-poor systems was
mostly based on the results of the distance determinations with the ISB technique. We
have shown in this paper that there is a high probability that
these conclusions were flawed, due the incorrect theoretical calibration
of the canonical p-factor law. If substantiated by future theoretical and improved empirical work,
this would be obviously very good news for the use of the Cepheid PL
relation as a primary distance indicator and would eliminate a strong concern which has been with us
for a number of years.

Regarding the slopes of the near-infrared Cepheid PL relations, particularly the K-band relation
 which is potentially the most accurate means
to determine Cepheid-based distances to galaxies due to its insensitivity to absorption
corrections and its small intrinsic dispersion as compared to optical PL relations
(as recently impressively demonstrated by the work of Persson et al. 2004), we can now say
that the present study suggests that
the Milky Way relation agrees in slope with the observed LMC relation to within 1 $\sigma$,
providing evidence for no change of the slope in the metallicity regime from -0.3 dex
to solar either. Very recent empirical results on the Cepheid K-band PL relation in NGC300 (Gieren
et al. 2005) do also indicate excellent agreement with the LMC PL relation
in K of Persson et al., strengthening the evidence that the slope of the K-band PL relation 
is metallicity-independent as well. However, the K-band PL relation has yet to be studied
for more metal-poor galaxies to confirm this.

\acknowledgments
WG and  GP  gratefully acknowledge 
financial support for this
work from the Chilean Center for Astrophysics FONDAP 15010003. 
Support from the Polish KBN grant No 2P03D02123 and BST grant for 
Warsaw University Observatory is also acknowledged. PF acknowledges Gustav Tammann
for sending us the most updated ZAMS distance moduli of Galactic cluster Cepheids.
We thank Roeland van der Marel for sending us his computer program for the
calculation of the tilt corrections to the LMC distance moduli.

\begin{deluxetable}{ccccccccc}
\tablewidth{0pc}
\tablecaption{ISB Distance and Radius Solutions for LMC Cepheids}
\tablehead{
\colhead{Cepheid}  & \colhead{log P}  & \colhead{$(m-M)_{0}$} &
\colhead{$ \sigma_{(m-M)}$} & \colhead{$\Delta m$} & 
\colhead{$(m-M)_{0, LMC}$} & \colhead{R} & \colhead{$\sigma_{R}$}
& \colhead{$\Delta_{\phi} $}\\
\colhead{}  & \colhead{[days]}  & \colhead{[mag]} &
\colhead{[mag]} & \colhead{[mag]} & \colhead{[mag]} & \colhead{[${R}_{\odot}$]} &
\colhead{[${R}_{\odot}$]} & \colhead{}\\
}
\startdata
HV12199 &  0.421469 & 18.336 & 0.094 & -0.058 & 18.394 & 25.0 & 1.1 & 0.025 \\ 
HV12203 &  0.470427 & 18.481 & 0.092 & -0.059 & 18.540 & 28.3 & 1.2 & 0.050 \\ 
HV12202 &  0.491519 & 18.289 & 0.072 & -0.059 & 18.348 & 28.5 & 1.0 & 0.025 \\ 
HV12197 &  0.497456 & 18.165 & 0.058 & -0.058 & 18.223 & 25.9 & 0.7 & -0.020 \\ 
HV12204 &  0.536402 & 18.202 & 0.044 & -0.059 & 18.261 & 28.3 & 0.6 & 0.010 \\ 
HV12198 &  0.546887 & 18.314 & 0.028 & -0.059 & 18.373 & 29.8 & 0.4 & 0.015 \\ 
HV12816 &  0.959466 & 18.328 & 0.087 & -0.076 & 18.404 & 54.1 & 2.2 & 0.035 \\ 
HV12815 &  1.416910 & 18.296 & 0.028 & -0.075 & 18.371 & 126.3 & 1.6 & -0.025\\ 
HV899 &  1.492040 & 18.769 & 0.013 & 0.017 & 18.752 & 160.2 & 0.9 & 0.030 \\ 
HV879 &  1.566170 & 18.532 & 0.040 & 0.044 & 18.488 & 163.3 & 3.0 & 0.025 \\ 
HV909 &  1.574990 & 18.397 & 0.029 & 0.048 & 18.349 & 155.1 & 2.1 & -0.055 \\ 
HV2257 &  1.595150 & 18.788 & 0.028 & 0.054 & 18.734 & 197.7 & 2.6 & 0.010 \\ 
HV2338 &  1.625350 & 18.663 & 0.023 & 0.070 & 18.593 & 199.4 & 2.2 & -0.005 \\ 
\enddata
\end{deluxetable}

\begin{deluxetable}{ccccc}
\tablewidth{0pc}
\tablecaption{Individual Radial Velocity Observations of HV12816}
\tablehead{
\colhead{HJD-2400000}  & \colhead{Vr}  & \colhead{$\sigma_{\rm Vr}$}\\
}
\startdata
 51163.699 &   270.88 &    0.30 \\ 
 51164.672 &   271.81 &    0.30 \\ 
 51165.711 &   276.32 &    0.30 \\ 
 51167.719 &   279.12 &    0.30 \\ 
 51168.699 &   286.94 &    0.30 \\ 
 51169.566 &   293.09 &    0.30 \\ 
 51170.586 &   294.08 &    0.30 \\ 
 51171.590 &   282.75 &    0.30 \\ 
 51172.703 &   270.87 &    0.30 \\ 
 51174.598 &   275.52 &    0.30 \\ 
 51175.844 &   275.69 &    0.30 \\ 
 51176.582 &   278.32 &    0.30 \\ 
 51178.836 &   293.81 &    0.30 \\ 
 51179.832 &   294.01 &    0.30 \\ 
 51181.633 &   271.19 &    0.30 \\ 
 51182.555 &   271.41 &    0.30 \\ 
 51193.762 &   275.97 &    0.30 \\ 
 51545.797 &   271.72 &    0.27 \\ 
 51546.785 &   271.47 &    0.13 \\ 
 51547.746 &   273.84 &    0.09 \\ 
 51548.773 &   277.06 &    0.17 \\ 
 51548.781 &   276.33 &    0.14 \\ 
 51549.770 &   276.36 &    0.07 \\ 
 51550.703 &   282.53 &    0.08 \\ 
 51551.770 &   290.72 &    0.08 \\ 
 51552.754 &   294.74 &    0.17 \\ 
 51564.738 &   270.77 &    0.14 \\ 
 51565.602 &   272.42 &    0.10 \\ 
 51566.703 &   276.93 &    0.09 \\ 
 51567.672 &   275.20 &    0.08 \\ 
 51568.594 &   280.05 &    0.08 \\ 
 51569.641 &   287.95 &    0.08 \\ 
 51570.707 &   294.64 &    0.14 \\ 
\enddata
\end{deluxetable}

\setcounter{table}{1}

\begin{deluxetable}{ccccc}
\tablewidth{0pc}
\tablecaption{Concluded}
\tablehead{
\colhead{HJD-2400000}  & \colhead{Vr}  & \colhead{$\sigma_{\rm Vr}$}\\
}
\startdata
 51571.664 &   293.19 &    0.21 \\ 
 51854.660 &   284.18 &    0.38 \\ 
 51909.762 &   276.43 &    0.41 \\ 
 52264.832 &   279.61 &    0.27 \\ 
 52270.746 &   286.08 &    0.24 \\ 
 52595.801 &   276.83 &    0.12 \\ 
 52601.699 &   282.88 &    0.34 \\ 
 52601.758 &   281.23 &    0.30 \\ 
 52603.758 &   271.32 &    0.22 \\ 
\enddata
\end{deluxetable}

\begin{deluxetable}{ccccccccc}
\tablewidth{0pc}
\tablecaption{Absolute Magnitudes of LMC Cepheids}
\tablehead{
\colhead{Cepheid} & \colhead{log P} & \colhead{$<M_{V}>$}  
& \colhead{$<M_{I}> $} & \colhead{$<M_{J}>$} & \colhead{$<M_{K}>$} 
& \colhead{$<M_{W}>$} & \colhead{E(B-V)} & \colhead{[Fe/H]}\\
\colhead{} & \colhead{[days]} & \colhead{[mag]}  & \colhead{[mag]} &
\colhead{[mag]} & \colhead{[mag]}  & \colhead{[mag]} & \colhead{[mag]} &
\colhead{[dex]}\\
}
\startdata
HV12199 &  0.421469 & -2.269 & -2.870 & -3.075 & -3.672 & -3.777 & 0.060 & -0.50 \\
HV12203 &  0.470427 & -2.552 & -3.153 & -3.562 & -3.930 & -4.060 & 0.060 & -0.50 \\ 
HV12202 &  0.491519 & -2.425 & -3.050 & -3.553 & -3.922 & -3.992 & 0.060 & -0.50 \\ 
HV12197 &  0.497456 & -2.273 & -2.918 & -3.338 & -3.728 & -3.891 & 0.060 & -0.50 \\ 
HV12204 &  0.536402 & -2.702 & -3.239 & -3.626 & -3.981 & -4.050 & 0.060 & -0.50 \\ 
HV12198 &  0.546887 & -2.565 & -3.202 & -3.675 & -4.030 & -4.165 & 0.060 & -0.50 \\ 
HV12816 &  0.959466 & -4.044 & -4.630 & -5.028 & -5.366 & -5.514 & 0.070 & -- \\ 
HV12815 &  1.416910 & -5.044 & -5.899 & -6.500 & -6.988 & -7.190 & 0.070 & -- \\ 
HV899 &  1.492040 & -5.734 & -6.543 & -7.097 & -7.541 & -7.763 & 0.110 & -- \\ 
HV879 &  1.566170 & -5.365 & -6.320 & -6.989 & -7.521 & -7.763 & 0.060 & -0.55 \\ 
HV909 &  1.574990 & -5.835 & -6.597 & -7.107 & -7.524 & -7.747 & 0.058 & -0.27 \\ 
HV2257 &  1.595150 & -5.961 & -6.860 & -7.475 & -7.959 & -8.216 & 0.060 & -0.36 \\ 
HV2338 &  1.625350 & -6.040 & -6.922 & -7.514 & -7.981 & -8.254 & 0.040 & -0.37 \\ 
\enddata
\end{deluxetable}

\begin{deluxetable}{ccccccc}
\tablewidth{0pc}
\tablecaption{Slopes of the Period-Luminosity Relation from the ISB
Technique assuming the Canonical p-factor Law}
\tablehead{
\colhead{Band}  & \colhead{LMC}  & \colhead{$\sigma$} & \colhead{Milky
Way} & \colhead{$\sigma$} & \colhead{LMC(OGLE-II/Persson et al.)} & \colhead{$\sigma$}  \\
}
\startdata
V &  -3.048 & 0.093 & -3.082 & 0.133 & -2.775 & 0.031 \\
I &  -3.289 & 0.079 & -3.312 & 0.109 & -2.977 & 0.021 \\
W &  -3.650 & 0.074 & -3.660 & 0.100 & -3.300 & 0.011 \\
J &  -3.476 & 0.078 & -3.510 & 0.095 & -3.153 & 0.051 \\
K &  -3.540 & 0.072 & -3.639 & 0.097 & -3.261 & 0.042 \\
\enddata
\end{deluxetable}

\begin{deluxetable}{c c c c c c c c}
\tablewidth{0pc}
\tablecaption{Galactic Cluster Cepheid ZAMS-fitting and ISB Distance
Moduli, for Canonical and Revised p-factor Laws}
\tablehead{
\colhead{Cepheid} & \colhead{$\log P$} & \colhead{$(m-M)_{0,ZAMS}$} &
\colhead{$\sigma$} & \colhead{$\Delta (m-M)_{0,old}$} & \colhead{$\sigma$} & 
\colhead{$\Delta (m-M)_{0,new}$} & \colhead{$\sigma$} \\
}
\startdata
   CVMon &  0.730685 &  11.21 &  0.04 & $ -0.21$ &  0.05 & $ -0.05$ & 
0.05\\
    VCen &  0.739882 &   9.17 &  0.04 & $  0.01$ &  0.07 & $  0.16$ & 
0.07\\
   CSVel &  0.771201 &  12.59 &  0.14 & $ -0.17$ &  0.15 & $ -0.02$ & 
0.15\\
    USgr &  0.828997 &   9.05 &  0.10 & $ -0.21$ &  0.10 & $ -0.07$ & 
0.10\\
    SNor &  0.989194 &   9.84 &  0.04 & $  0.07$ &  0.05 & $  0.18$ & 
0.05\\
 V340Nor &  1.052579 &  11.19 &  0.11 & $ -0.04$ &  0.21 & $  0.06$ & 
0.21\\
    XCyg &  1.214482 &  10.30 &  0.05 & $  0.12$ &  0.05 & $  0.19$ & 
0.05\\
   VYCar &  1.276818 &  11.62 &  0.09 & $ -0.12$ &  0.09 & $ -0.06$ & 
0.09\\
   RZVel &  1.309564 &  11.23 &  0.30 & $ -0.21$ &  0.30 & $ -0.16$ & 
0.30\\
   WZSgr &  1.339443 &  11.27 &  0.04 & $  0.02$ &  0.06 & $  0.06$ & 
0.06\\
   SWVel &  1.370016 &  12.06 &  0.05 & $ -0.07$ &  0.06 & $ -0.02$ & 
0.06\\
    TMon &  1.431915 &  11.10 &  0.14 & $ -0.29$ &  0.15 & $ -0.26$ & 
0.15\\
\enddata
\end{deluxetable}
                                                                        
\begin{deluxetable}{c c c c c c}
\tablewidth{0pc}
\tablecaption{ISB LMC Cepheid Distance Moduli assuming the Revised
p-factor Law}
\tablehead{
\colhead{Cepheid} & \colhead{$\log P$} & \colhead{$(m-M)_0$} & \colhead{$\sigma(m-M)$} &
\colhead{$\Delta m$} & \colhead{$(m-M)_{0,LMC}$} \\
\colhead{ (1)} & \colhead{(2)} & \colhead{(3)} & \colhead{(4)} 
& \colhead{(5)} & \colhead{(6)} \\
}
\startdata 
   HV12199 & 0.421469 & 18.545 & 0.094 & $-0.058$ & 18.603\\
   HV12203 & 0.470427 & 18.683 & 0.092 & $-0.059$ & 18.742\\
   HV12202 & 0.491519 & 18.486 & 0.072 & $-0.059$ & 18.545\\
   HV12197 & 0.497456 & 18.362 & 0.058 & $-0.058$ & 18.420\\
   HV12204 & 0.536402 & 18.392 & 0.044 & $-0.059$ & 18.451\\
   HV12198 & 0.546887 & 18.502 & 0.028 & $-0.059$ & 18.561\\
   HV12816 & 0.959466 & 18.444 & 0.087 & $-0.076$ & 18.520\\
   HV12815 & 1.416910 & 18.328 & 0.028 & $-0.075$ & 18.403\\
     HV899 & 1.492040 & 18.786 & 0.013 & $ 0.017$ & 18.769\\
     HV879 & 1.566170 & 18.535 & 0.040 & $ 0.044$ & 18.491\\
     HV909 & 1.574990 & 18.398 & 0.029 & $ 0.048$ & 18.350\\
    HV2257 & 1.595150 & 18.786 & 0.028 & $ 0.054$ & 18.732\\
    HV2338 & 1.625350 & 18.654 & 0.023 & $ 0.070$ & 18.584\\
\enddata
\end{deluxetable}

\begin{deluxetable}{ccccccccccccccc}
\tabletypesize{\scriptsize}
\rotate
\tablewidth{0pc}
\tablecaption{Revised Galactic Cepheid ISB Distances, Radii and Absolute Magnitudes
assuming the Revised p-factor Law}
\tablehead{
\colhead{Cepheid} & \colhead{log P} & \colhead{$(m-M)_0$} & \colhead{$\sigma_{(m-M)}$} &
\colhead{$R$} & \colhead{$\sigma_R$} & \colhead{$M_B$} & \colhead{$M_V$} & \colhead{$M_I$}
& \colhead{$M_J$} & \colhead{$M_H$} & \colhead{$M_K$} & \colhead{$M_W$} &
\colhead{$E(B-V)$} & \colhead{$\Delta \phi$} \\
}
\startdata
   SU Cas &   0.289884 &  8.399 & 0.070 &  32.7 &  1.0 & $-2.960$ & $-3.375$ & $-3.876$ & $-4.160$ & $-4.347$ & $-4.372$ & $-4.632$ &  0.287 & $ 0.000$\\
   EV Sct &   0.490098 & 11.448 & 0.105 &  37.4 &  1.8 & $-3.078$ & $-3.543$ & $-4.177$ & $-4.429$ & $-4.617$ & $-4.640$ & $-5.133$ &  0.679 & $ 0.045$\\
   BF Oph &   0.609329 &  9.448 & 0.034 &  34.7 &  0.5 & $-2.312$ & $-2.928$ & $-3.576$ & $-4.021$ & $-4.290$ & $-4.359$ & $-4.554$ &  0.247 & $ 0.035$\\
    T Vel &   0.666501 &  9.970 & 0.060 &  36.3 &  1.0 & $-2.216$ & $-2.860$ & $-3.537$ & $-4.055$ & $-4.347$ & $-4.428$ & $-4.559$ &  0.281 & $ 0.000$\\
   $\delta$ Cep &   0.729678 &  7.242 & 0.044 &  45.1 &  0.9 & $-3.027$ & $-3.588$ & $-4.217$ & $-4.628$ & $-4.909$ & $-4.965$ & $-5.166$ &  0.092 & $ 0.000$\\
   CV Mon &   0.730685 & 11.159 & 0.034 &  43.7 &  0.7 & $-2.627$ & $-3.207$ & $-3.965$ & $-4.434$ & $-4.718$ & $-4.819$ & $-5.110$ &  0.714 & $ 0.015$\\
    V Cen &   0.739882 &  9.330 & 0.063 &  45.1 &  1.3 & $-2.869$ & $-3.450$ & $-4.112$ & $-4.569$ & $-4.849$ & $-4.925$ & $-5.111$ &  0.289 & $ 0.000$\\
   CS Vel &   0.771201 & 12.567 & 0.064 &  41.3 &  1.2 & $-3.161$ & $-3.661$ & $-4.185$ & $-4.539$ & $-4.766$ & $-4.827$ & $-4.977$ &  0.847 & $ 0.000$\\
   BB Sgr &   0.821971 &  9.660 & 0.028 &  53.2 &  0.7 & $-2.958$ & $-3.659$ & $-4.403$ & $-4.865$ & $-5.170$ & $-5.243$ & $-5.527$ &  0.284 & $-0.035$\\
    U Sgr &   0.828997 &  8.977 & 0.021 &  50.8 &  0.5 & $-2.925$ & $-3.617$ & $-4.353$ & $-4.808$ & $-5.090$ & $-5.163$ & $-5.463$ &  0.403 & $ 0.000$\\
  $\eta$ Aql &   0.855930 &  7.125 & 0.045 &  51.5 &  1.1 & $-3.081$ & $-3.716$ & $-4.406$ & $-4.852$ & $-5.147$ & $-5.208$ & $-5.448$ &  0.149 & $ 0.000$\\
    S Sge &   0.923352 &  9.334 & 0.035 &  62.5 &  1.0 & $-3.456$ & $-4.136$ & $-4.812$ & $-5.276$ & $-5.566$ & $-5.628$ & $-5.834$ &  0.127 & $ 0.000$\\
    S Nor &   0.989194 & 10.020 & 0.032 &  74.4 &  1.1 & $-3.457$ & $-4.213$ & $-4.968$ & $-5.523$ & $-5.846$ & $-5.930$ & $-6.108$ &  0.189 & $ 0.000$\\
    Z Lac &   1.036854 & 11.549 & 0.043 &  74.8 &  1.5 & $-3.773$ & $-4.468$ & $-5.205$ & $-5.625$ & $-5.939$ & $-6.002$ & $-6.319$ &  0.404 & $ 0.000$\\
   XX Cen &   1.039548 & 11.216 & 0.022 &  72.7 &  0.8 & $-3.532$ & $-4.256$ & $-4.999$ & $-5.517$ & $-5.821$ & $-5.904$ & $-6.121$ &  0.260 & $-0.040$\\
 V340 Nor &   1.052579 & 11.246 & 0.185 &  70.3 &  6.0 & $-3.085$ & $-3.914$ & $-4.779$ & $-5.326$ & $-5.678$ & $-5.770$ & $-6.085$ &  0.315 & $ 0.000$\\
   UU Mus &   1.065819 & 12.687 & 0.084 &  77.4 &  3.0 & $-3.526$ & $-4.256$ & $-5.022$ & $-5.592$ & $-5.910$ & $-6.002$ & $-6.178$ &  0.413 & $-0.005$\\
    U Nor &   1.101875 & 10.806 & 0.060 &  79.5 &  2.2 & $-3.804$ & $-4.506$ & $-5.232$ & $-5.735$ & $-6.018$ & $-6.108$ & $-6.329$ &  0.892 & $ 0.000$\\
   BN Pup &   1.135867 & 13.035 & 0.050 &  86.6 &  2.0 & $-3.848$ & $-4.597$ & $-5.353$ & $-5.864$ & $-6.184$ & $-6.266$ & $-6.495$ &  0.438 & $ 0.000$\\
   LS Pup &   1.150646 & 13.636 & 0.056 &  93.6 &  2.4 & $-4.008$ & $-4.767$ & $-5.515$ & $-6.035$ & $-6.363$ & $-6.439$ & $-6.643$ &  0.478 & $ 0.000$\\
   VW Cen &   1.177138 & 12.881 & 0.039 &  89.8 &  1.6 & $-3.224$ & $-4.114$ & $-5.005$ & $-5.707$ & $-6.099$ & $-6.213$ & $-6.351$ &  0.448 & $ 0.000$\\
    X Cyg &   1.214482 & 10.489 & 0.018 & 109.0 &  0.9 & $-4.192$ & $-5.060$ & $-5.837$ & $-6.342$ & $-6.684$ & $-6.760$ & $-7.010$ &  0.288 & $ 0.000$\\
    Y Oph &   1.233609 &  8.934 & 0.029 &  92.3 &  1.2 & $-4.215$ & $-4.925$ & $-5.718$ & $-6.132$ & $-6.394$ & $-6.458$ & $-6.916$ &  0.655 & $ 0.000$\\
   VY Car &   1.276818 & 11.556 & 0.022 & 115.8 &  1.2 & $-3.986$ & $-4.903$ & $-5.759$ & $-6.381$ & $-6.736$ & $-6.840$ & $-7.053$ &  0.243 & $-0.020$\\
   RY Sco &   1.307927 & 10.567 & 0.034 & 102.4 &  1.6 & $-4.447$ & $-5.113$ & $-5.859$ & $-6.322$ & $-6.593$ & $-6.676$ & $-6.985$ &  0.777 & $ 0.000$\\
   RZ Vel &   1.309564 & 11.073 & 0.029 & 117.6 &  1.6 & $-4.301$ & $-5.093$ & $-5.877$ & $-6.460$ & $-6.787$ & $-6.878$ & $-7.060$ &  0.335 & $ 0.000$\\
   WZ Sgr &   1.339443 & 11.334 & 0.047 & 124.4 &  2.7 & $-3.921$ & $-4.848$ & $-5.768$ & $-6.428$ & $-6.812$ & $-6.928$ & $-7.157$ &  0.467 & $ 0.000$\\
   WZ Car &   1.361977 & 12.961 & 0.066 & 114.3 &  3.5 & $-4.184$ & $-4.960$ & $-5.760$ & $-6.365$ & $-6.704$ & $-6.787$ & $-6.967$ &  0.384 & $ 0.000$\\
   VZ Pup &   1.364945 & 13.122 & 0.056 &  98.8 &  2.6 & $-4.362$ & $-5.050$ & $-5.762$ & $-6.231$ & $-6.533$ & $-6.594$ & $-6.836$ &  0.471 & $ 0.000$\\
   SW Vel &   1.370016 & 12.036 & 0.025 & 119.5 &  1.4 & $-4.252$ & $-5.060$ & $-5.885$ & $-6.485$ & $-6.827$ & $-6.931$ & $-7.132$ &  0.349 & $-0.020$\\
    T Mon &   1.431915 & 10.844 & 0.055 & 151.6 &  3.9 & $-4.432$ & $-5.401$ & $-6.276$ & $-6.921$ & $-7.303$ & $-7.404$ & $-7.598$ &  0.209 & $ 0.000$\\
   RY Vel &   1.449158 & 12.045 & 0.032 & 141.5 &  2.1 & $-4.719$ & $-5.527$ & $-6.328$ & $-6.911$ & $-7.209$ & $-7.303$ & $-7.538$ &  0.562 & $-0.005$\\
   AQ Pup &   1.478624 & 12.542 & 0.045 & 149.2 &  3.1 & $-4.669$ & $-5.533$ & $-6.427$ & $-6.969$ & $-7.321$ & $-7.423$ & $-7.778$ &  0.512 & $-0.055$\\
   KN Cen &   1.531857 & 13.134 & 0.045 & 186.7 &  3.9 & $-5.652$ & $-6.338$ & $-6.985$ & $-7.516$ & $-7.846$ & $-7.946$ & $-7.962$ &  0.926 & $ 0.005$\\
    l Car &   1.550816 &  8.749 & 0.022 & 179.5 &  1.8 & $-4.471$ & $-5.581$ & $-6.530$ & $-7.214$ & $-7.631$ & $-7.721$ & $-7.963$ &  0.170 & $-0.040$\\
    U Car &   1.588970 & 10.909 & 0.025 & 157.5 &  1.9 & $-4.658$ & $-5.551$ & $-6.416$ & $-7.038$ & $-7.387$ & $-7.488$ & $-7.722$ &  0.283 & $-0.050$\\
   RS Pup &   1.617420 & 11.556 & 0.064 & 207.5 &  6.1 & $-5.041$ & $-6.009$ & $-6.956$ & $-7.594$ & $-7.962$ & $-8.078$ & $-8.386$ &  0.446 & $ 0.000$\\
   SV Vul &   1.653162 & 12.088 & 0.037 & 222.6 &  3.8 & $-5.862$ & $-6.738$ & $-7.553$ & $-7.987$ & $-8.300$ & $-8.359$ & $-8.783$ &  0.570 & $-0.045$\\
\enddata
\end{deluxetable}

\begin{deluxetable}{ccccccc}
\tablewidth{0pc}
\tablecaption{ Slopes of the Period-Luminosity Relation from the ISB Technique
assuming the Revised p-factor Law}
\tablehead{\colhead{Band}  &  \colhead{LMC} & \colhead{$\sigma$} & \colhead{Milky
Way} &  \colhead{$\sigma$} & \colhead{LMC(OGLE-II/Persson et al.)} & \colhead{$\sigma$}\\
}
\startdata
 V  &  -2.867 &  0.093 &    -2.898 &  0.133 &   -2.775 &  0.031\\
 I  &  -3.108 &  0.079 &    -3.129 &  0.109 &   -2.977 &  0.021\\
 W  &  -3.469 &  0.074 &    -3.477 &  0.100 &   -3.300 &  0.011\\
 J  &  -3.295 &  0.078 &    -3.328 &  0.095 &   -3.153 &  0.051\\
 K  &  -3.359 &  0.072 &    -3.456 &  0.097 &   -3.261 &  0.042\\
\enddata
\end{deluxetable} 

\begin{deluxetable}{r c c c}
\tablewidth{0pc}
\tablecaption{\label{tab.dataref} References to the papers containing the
observational data for the four stars which have been added to the
sample of Storm et al. (2004) and the new radial velocity data
for $\ell$~Car.}
\tablehead{
\colhead{Star} & \colhead{Optical} & \colhead{Infra-red} & 
\colhead{Radial velocity} \\
 \colhead{}    & \colhead{photometry} & \colhead{photometry}
 & \colhead{}}
\startdata
CS~Vel & 1 & 2,3,4 & 5,6\\
S~Sge & 7,8,9 & 4,8 & 10\\
Z~Lac & 7,8 & 8 & 11,12\\
Y~Oph & 7,13,14  & 2,4 & 10 \\
$\ell$~Car & & & 15\\
\tablecomments{
1: Berdnikov et al. (1995),
2: Laney and Stobie (1992),
3: Schecter et al. (1992),
4: Welch et al. (1984),
5: Bersier et al. (1994),
6: Metzger et al. (1992),
7: Moffet and Barnes (1984),
8: Barnes et al. (1997),
9: Kiss (1998),
10: Gorynya et al. (1998),
11: Sugars and Evans (1996),
12: Imbert (1996)
13: Pel (1976),
14: Coulson and Caldwell (1985)
15: Taylor et al. (1997)
}
\enddata
\end{deluxetable}

\begin{figure}[htb] 
\vspace*{18cm}
\includegraphics{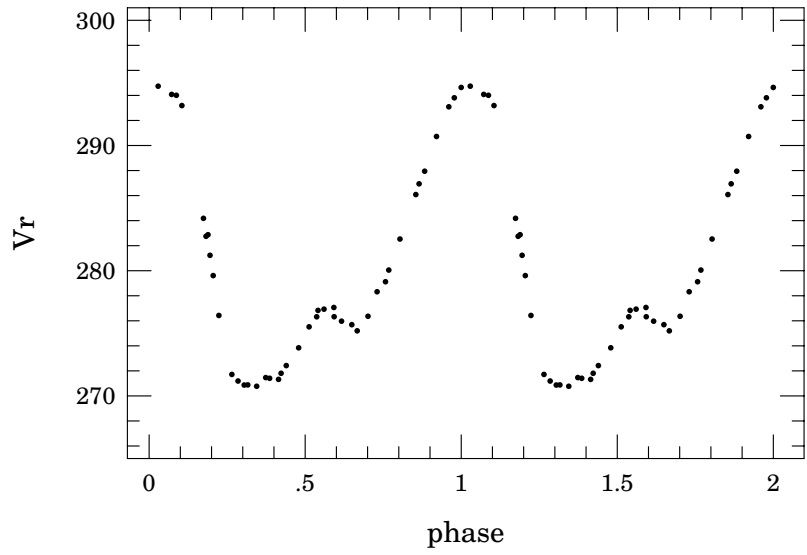} 
\caption{Radial velocity curve for the LMC Cepheid HV12816 from our new measurements in
Table 2. }
\end{figure}  

\clearpage

\begin{figure}[htb]  
\vspace*{18cm}
\includegraphics{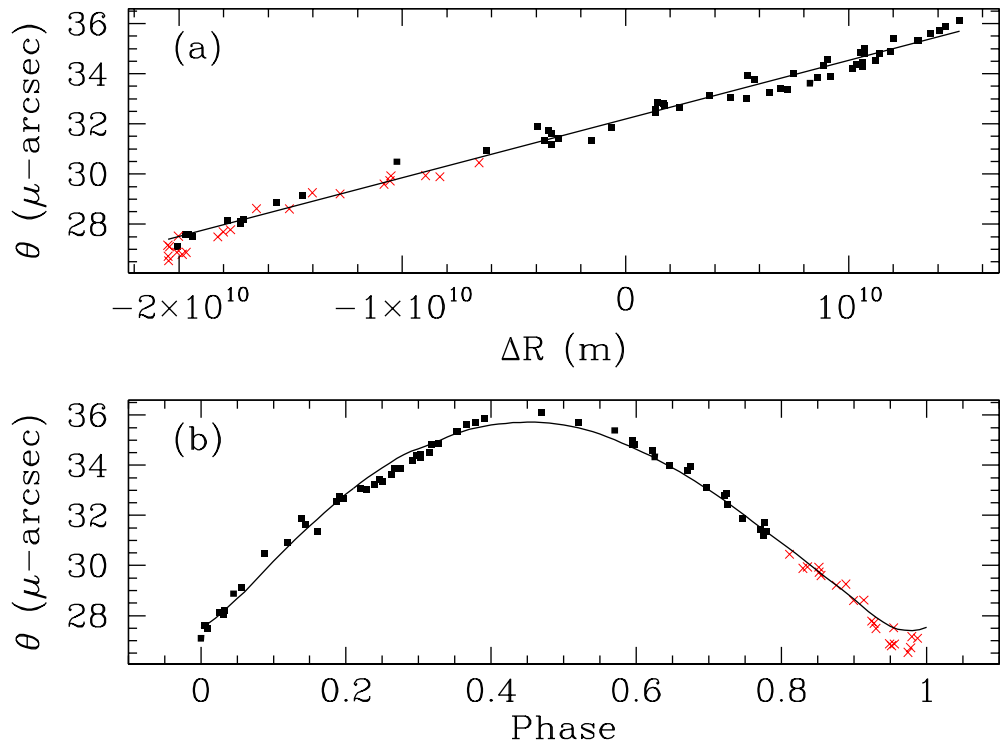}
\caption{Infrared surface brightness distance solution for the LMC Cepheid HV2257.
The points represent the photometrically determined angular diameters and the line in
panel (a) shows the bisector fit to the filled points. The curve in panel (b) delineates
the angular diameter curve obtained from integrating the radial velocity curve
of the star at the derived distance. Crosses in the 0.8-1.0 phase interval were
eliminated from the fit.  }
\end{figure}

\clearpage

\begin{figure}[htb]
\vspace*{18cm}
\includegraphics{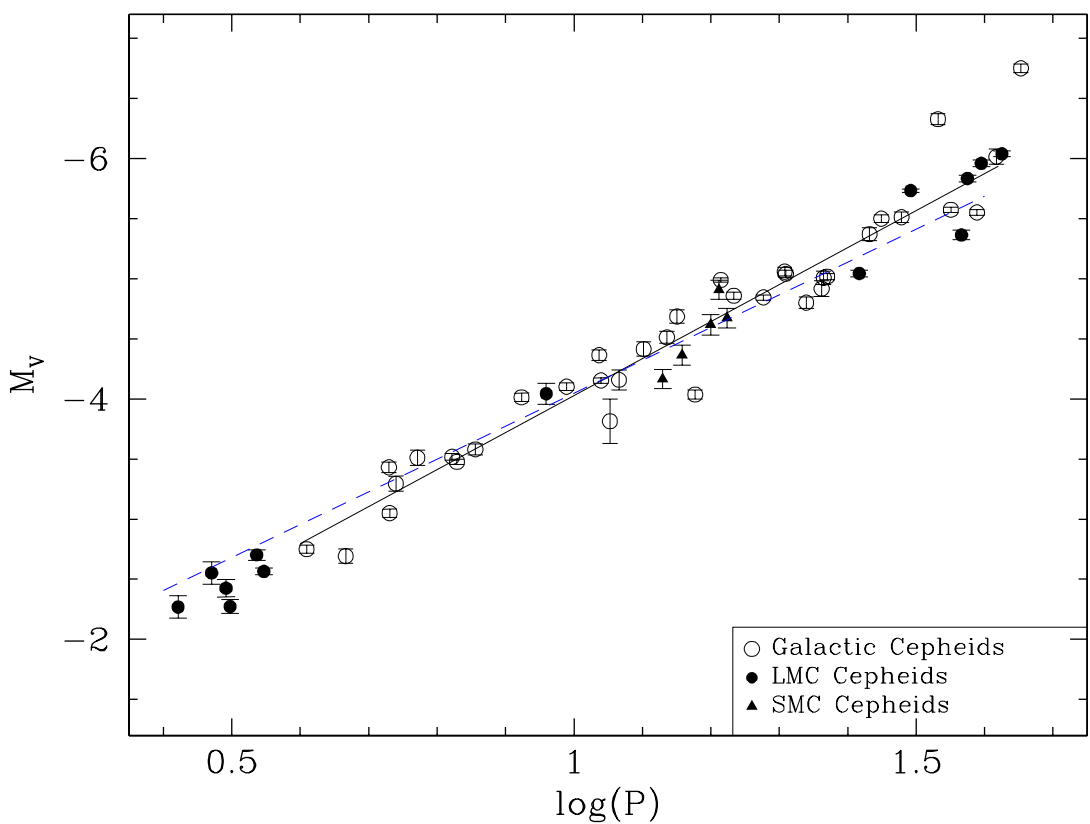}
\caption{The absolute V-band magnitudes derived from the ISB distances of 38 Milky Way,
13 LMC and 5 SMC Cepheids. The canonical p-factor relation was used for the calculation of
the ISB distances of all stars. The solid line is the best fit to the Galactic
data. The dashed line is the V-band PL relation in the LMC from the OGLE-II project, for an
assumed LMC distance modulus of 18.50. The ISB-based PL relation defined by the LMC Cepheids agrees
exceedingly well with the Milky Way ISB-based relation; both are significantly steeper
than the observed OGLE-II relation.}
\end{figure}

\clearpage

\begin{figure}[htb]
\vspace*{18cm}
\includegraphics{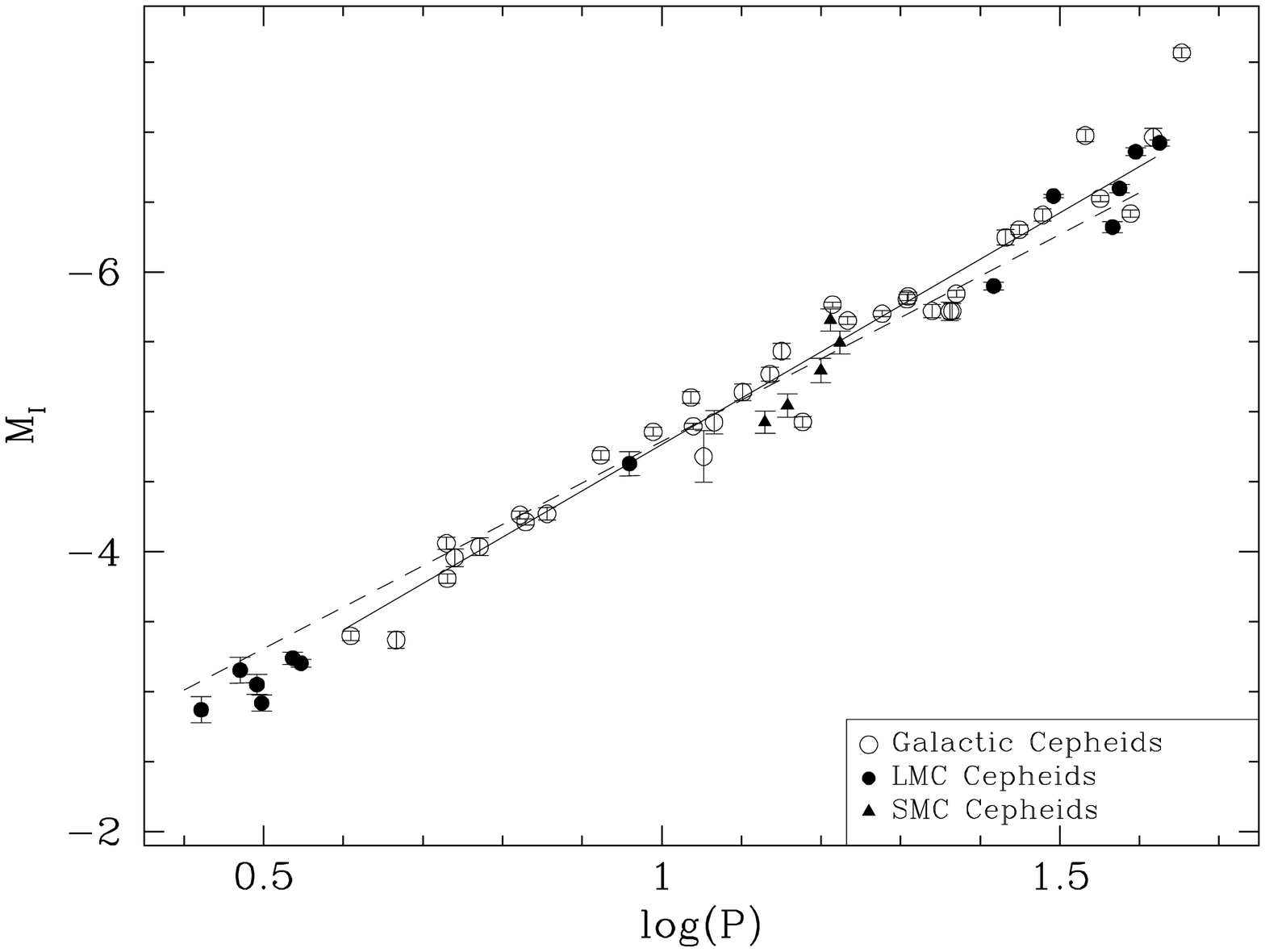}
\caption{Same as Fig. 3, for the absolute magnitudes in the I band.}
\end{figure}

\clearpage

\begin{figure}[htb]
\vspace*{18cm}
\includegraphics{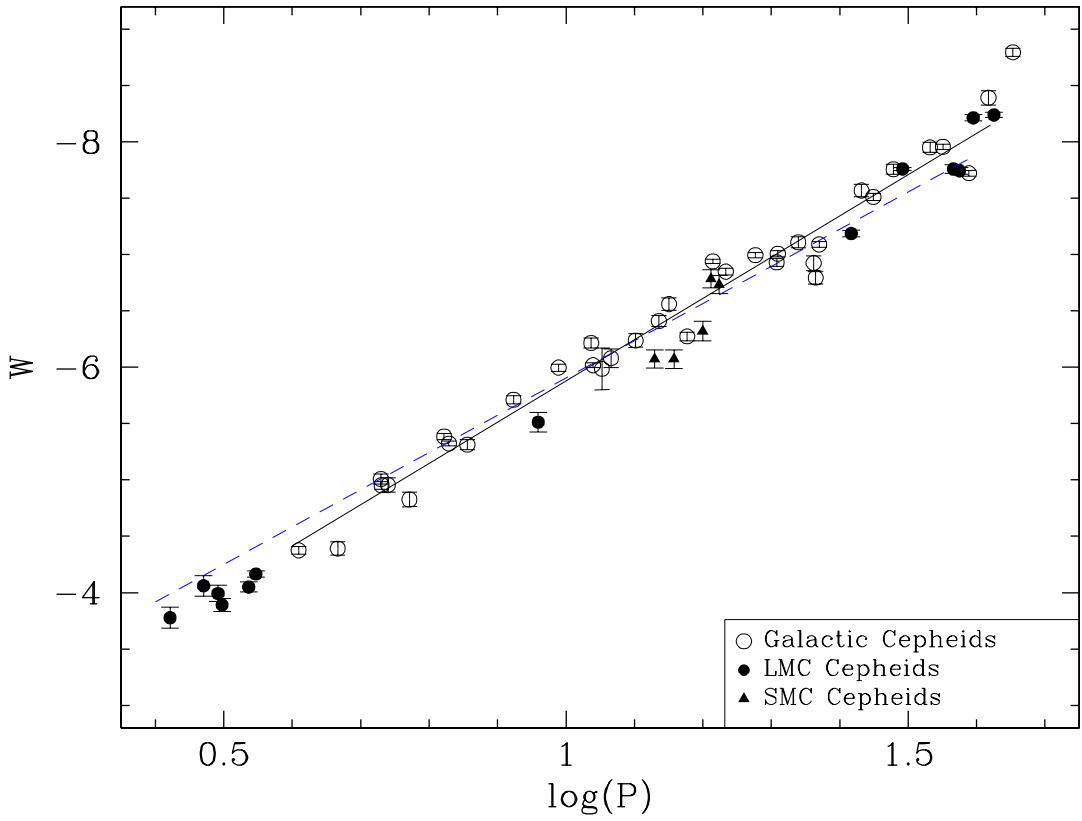}
\caption{Same as Fig. 3, for the absolute reddening-free W-band magnitudes.}
\end{figure}

\clearpage 

\begin{figure}[htb]
\vspace*{15cm}
\includegraphics{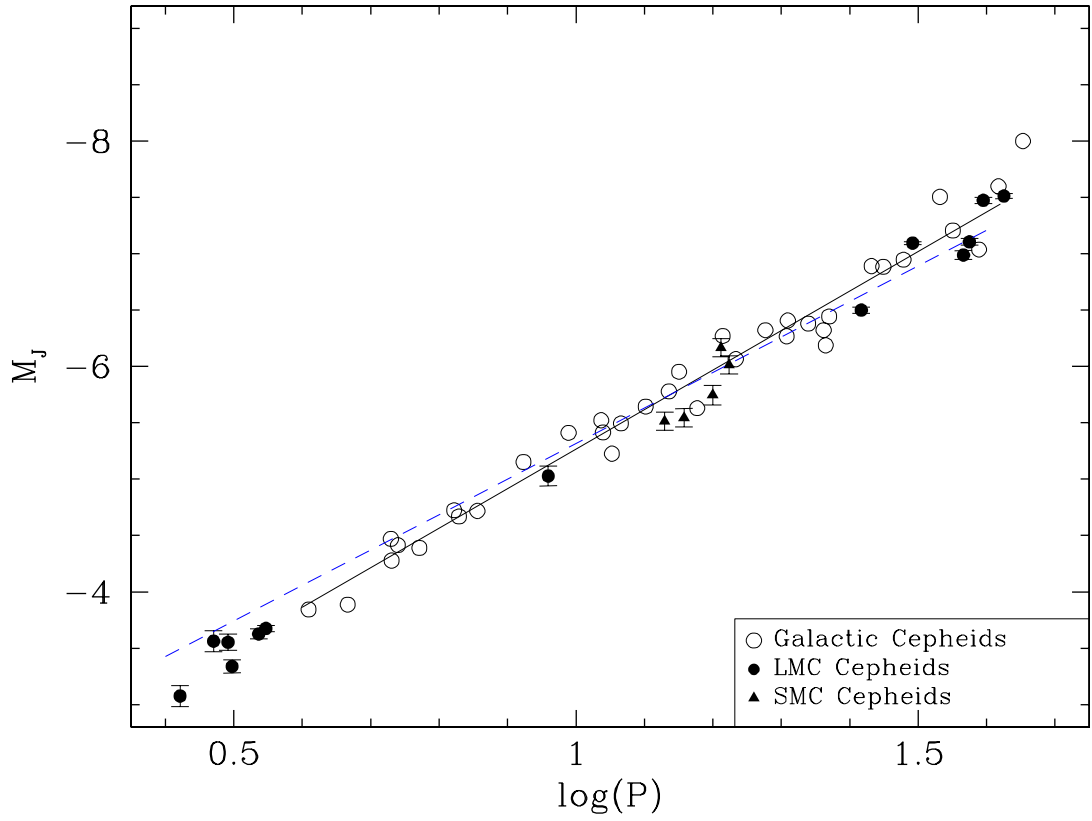}
\caption{Same as Fig. 3, for the absolute magnitudes in the J band. The dashed line
is the J-band PL relation in the LMC measured by Persson et al. (2004), for an
assumed LMC distance modulus of 18.50.}
\end{figure}

\clearpage

\begin{figure}[htb]
\vspace*{15cm}
\includegraphics{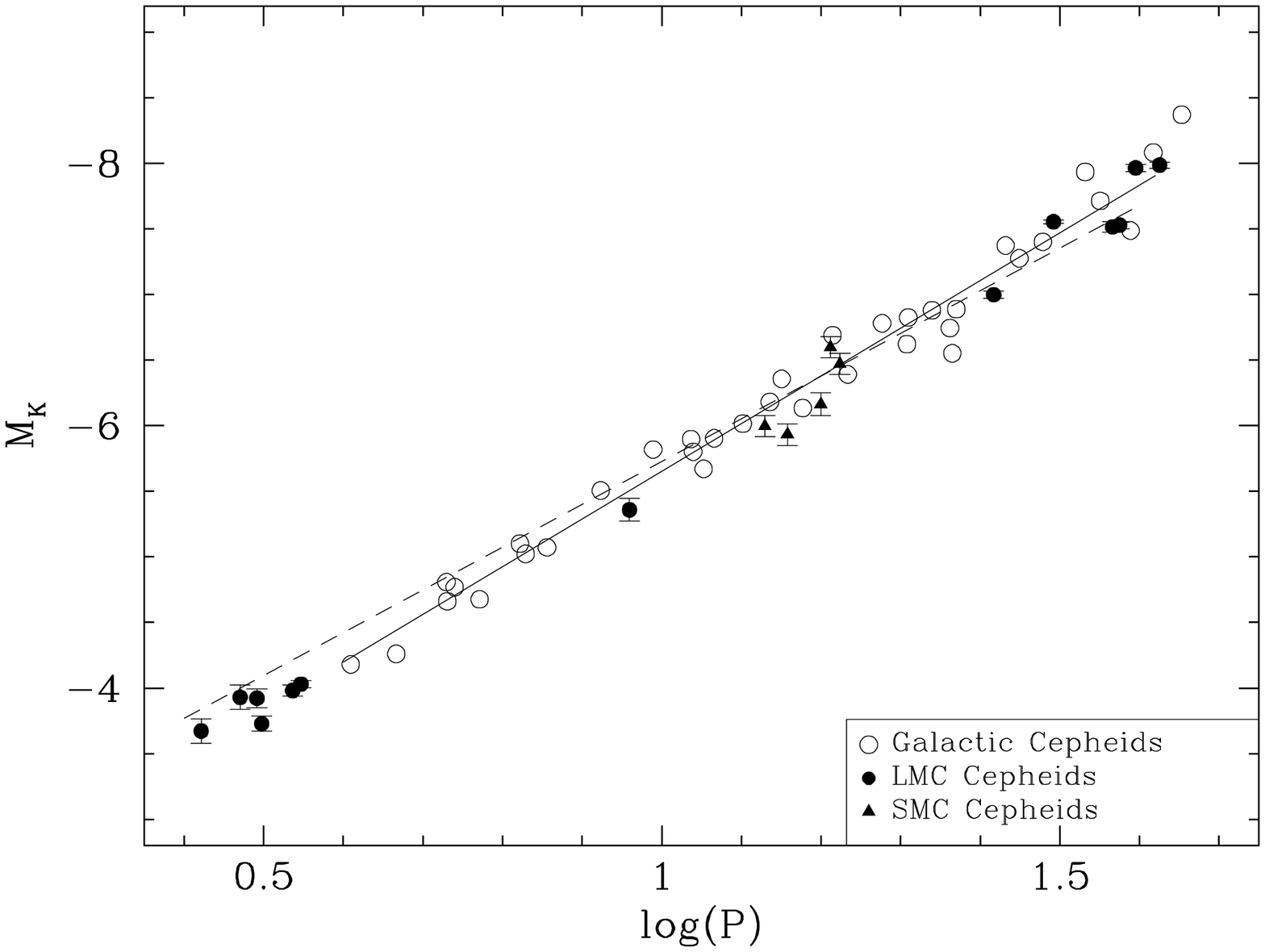}
\caption{Same as Fig. 6, for the absolute magnitudes in the K band.}
\end{figure}

\clearpage

\begin{figure}[htb]
\vspace*{15cm}
\includegraphics{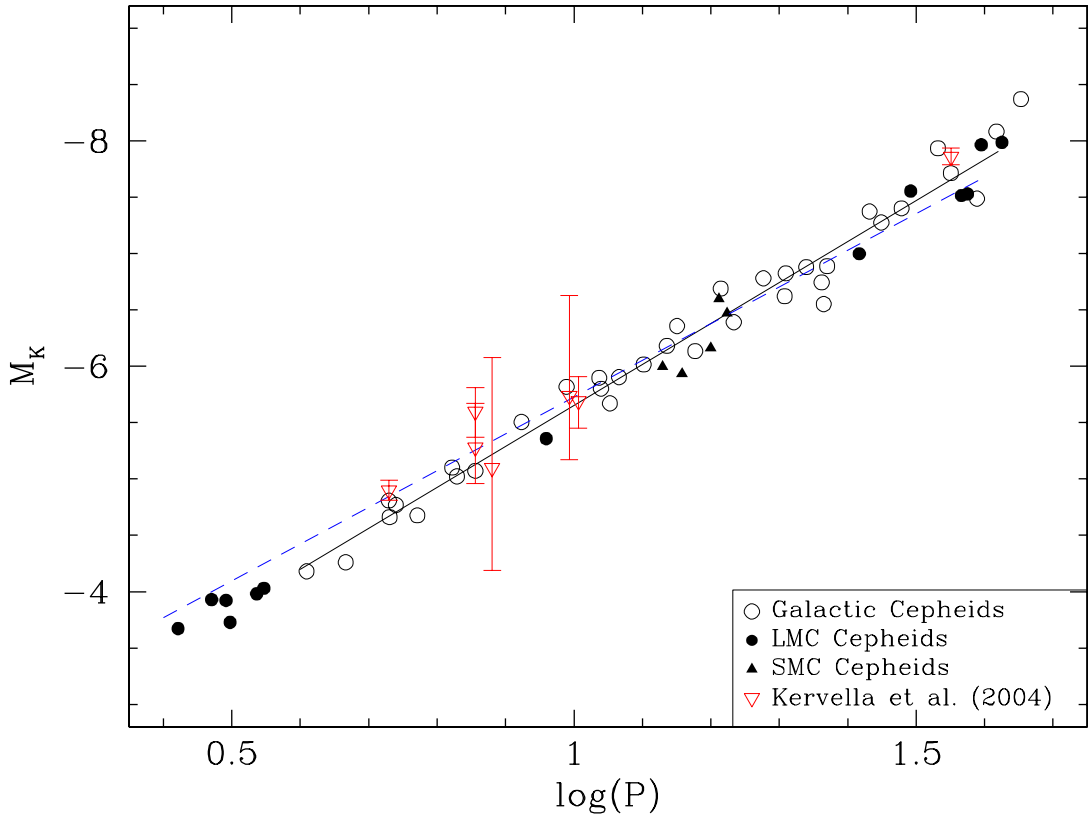}
\caption{Same as Fig. 7. The triangles are the absolute magnitudes of seven Milky Way
Cepheids whose angular diameter curves have been measured with the ESO VLTI by Kervella
et al. (2004b). Within the error bars, there is very good agreement with the 
absolute magnitudes of the Milky Way Cepheids whose angular diameters have been
determined with our adopted calibration of the surface brightness-color relation.}
\end{figure}

\clearpage

\begin{figure}[htb]
\vspace*{15cm}
\includegraphics{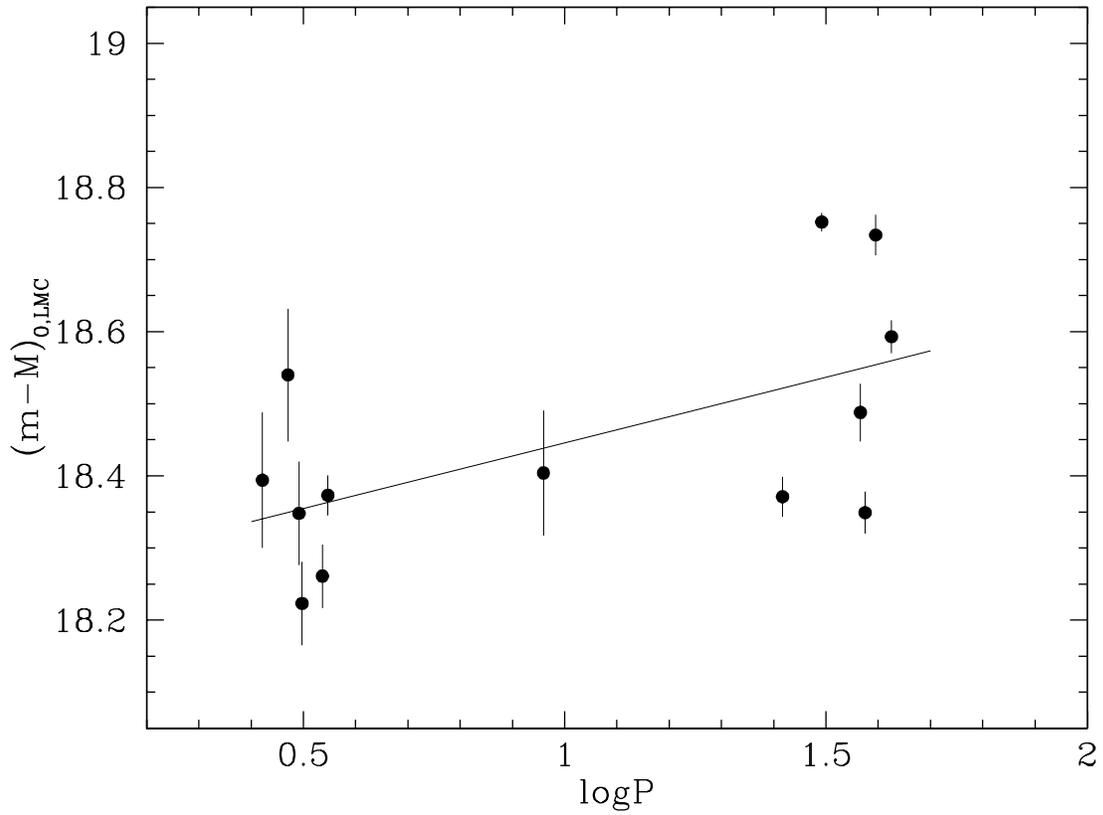}
\caption{The ISB-determined true distance moduli for LMC Cepheids, calculated with
the canonical p-factor law, plotted against
their period. The distance moduli have been corrected for the tilt of the LMC plane
with respect to the line of sight with the model of van der Marel and Cioni (2001).
There is a significant trend of the distance moduli with period.
}
\end{figure}

\clearpage

\begin{figure}[htb]
\vspace*{15cm}
\includegraphics{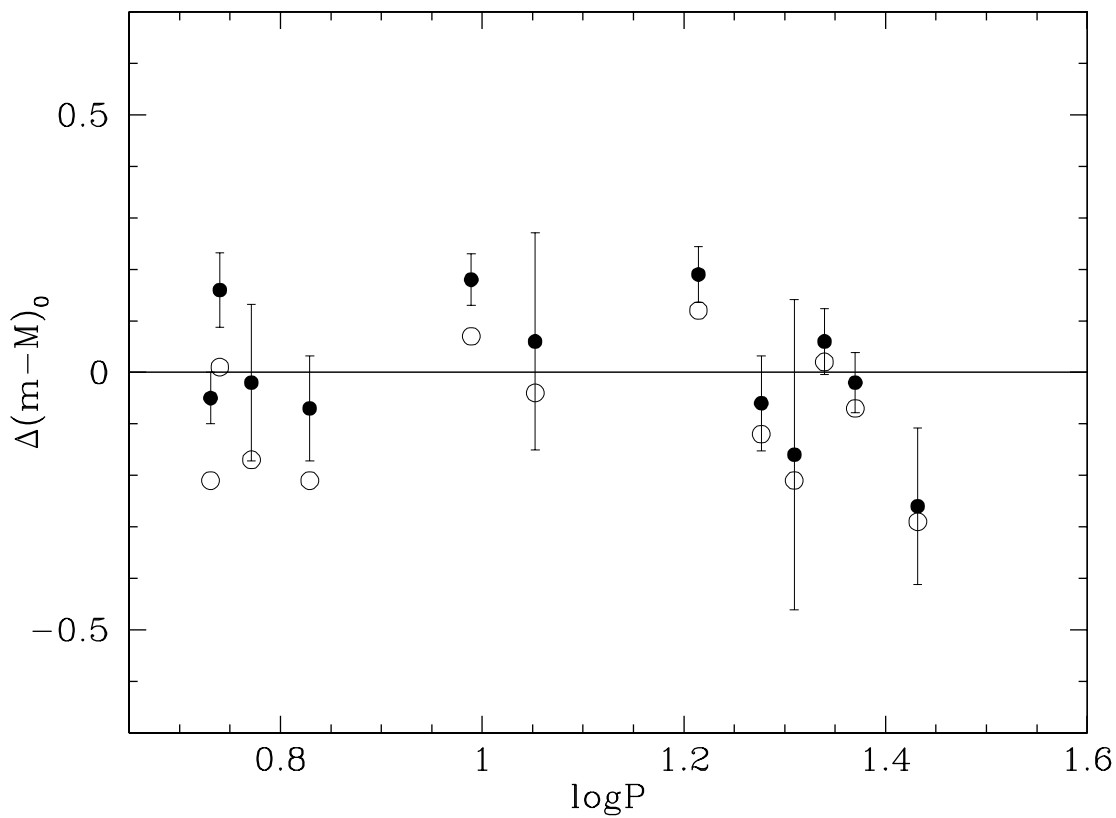}
\caption{The ISB- and ZAMS-fitting distance modulus difference, in the sense ISB-ZAMS,
for the twelve open cluster Cepheids in Table 5, plotted against their period. Open circles:
ISB distances have been calculated with the canonical p-factor law. Filled circles:
ISB distances have been calculated with the revised p-factor law of this paper. Error
bars have been plotted for the filled circles only and are the same for the open circles.
Using the canonical p-factor law, the mean difference ISB-ZAMS is -0.09 mag. Using the
revised p-factor law, the mean difference is zero.
}
\end{figure}

\clearpage

\begin{figure}[htb]
\vspace*{15cm}
\includegraphics{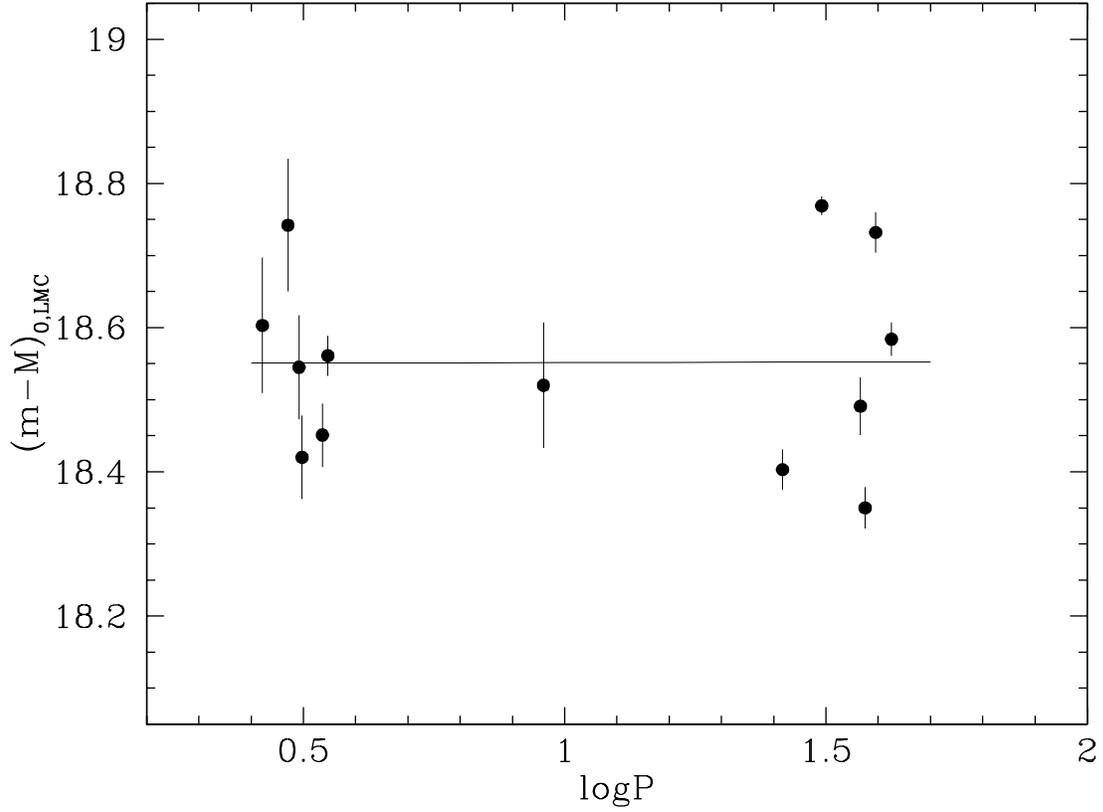}
\caption{The true distance moduli of the LMC Cepheids from their ISB distances, now 
calculated with the revised p-factor law derived in this paper. There is no trend
with period anymore. The LMC barycenter distance from these data is 18.56 $\pm$ 0.04 mag. 
}
\end{figure}

\clearpage

\begin{figure}[htb]
\vspace*{15cm}
\includegraphics{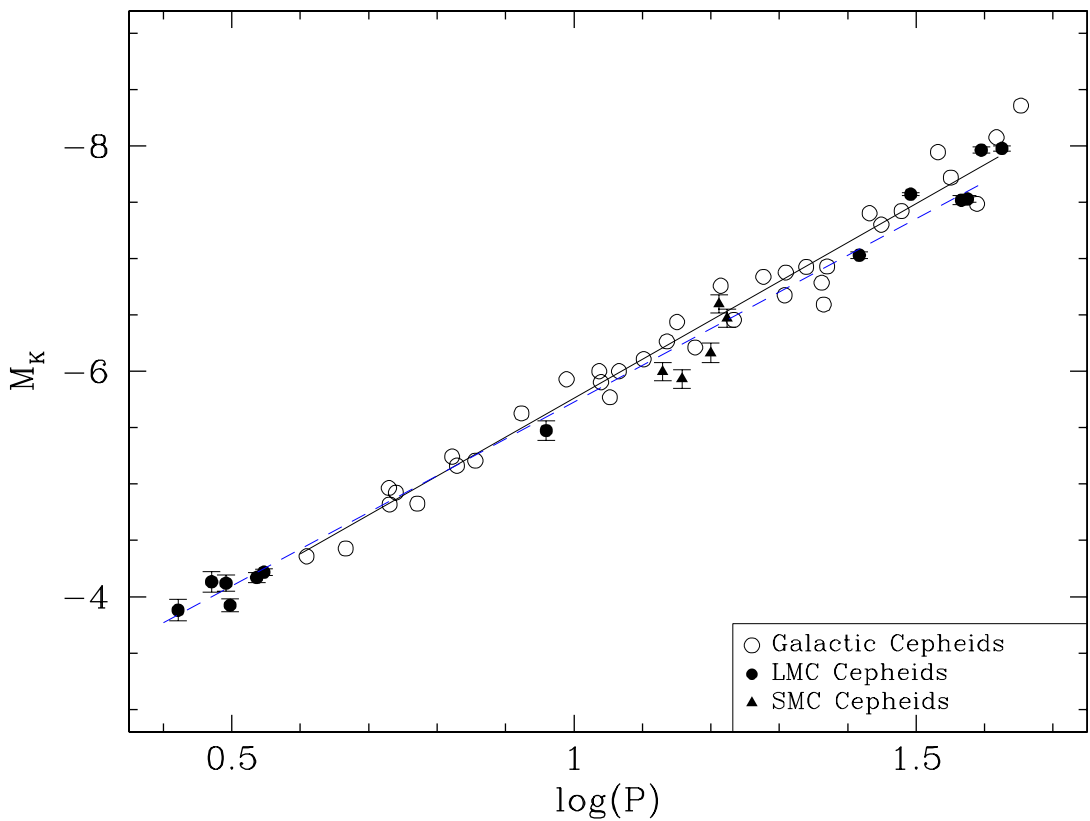}
\caption{The absolute magnitudes in the K band from the ISB distances of Milky Way, LMC
and SMC Cepheids, all calculated with the revised p-factor law of this paper.
The solid line is the best fit to the Milky Way Cepheid data. The dashed line is the
Persson et al. (2004) K-band PL relation for the LMC, for an assumed LMC distance modulus
of 18.50. The slopes of the PL relations for both the LMC and Milky Way derived from the ISB technique 
agree now very well with the slope of the observed K-band relation in the LMC. The slight mean
offset of the 5 SMC Cepheids from the Galactic relation has been used by Storm et al. (2004)
to constrain the metallicity effect on the zero point of the PL relation.
}
\end{figure}

\clearpage

\begin{figure}[htb]
\vspace*{15cm}
\includegraphics{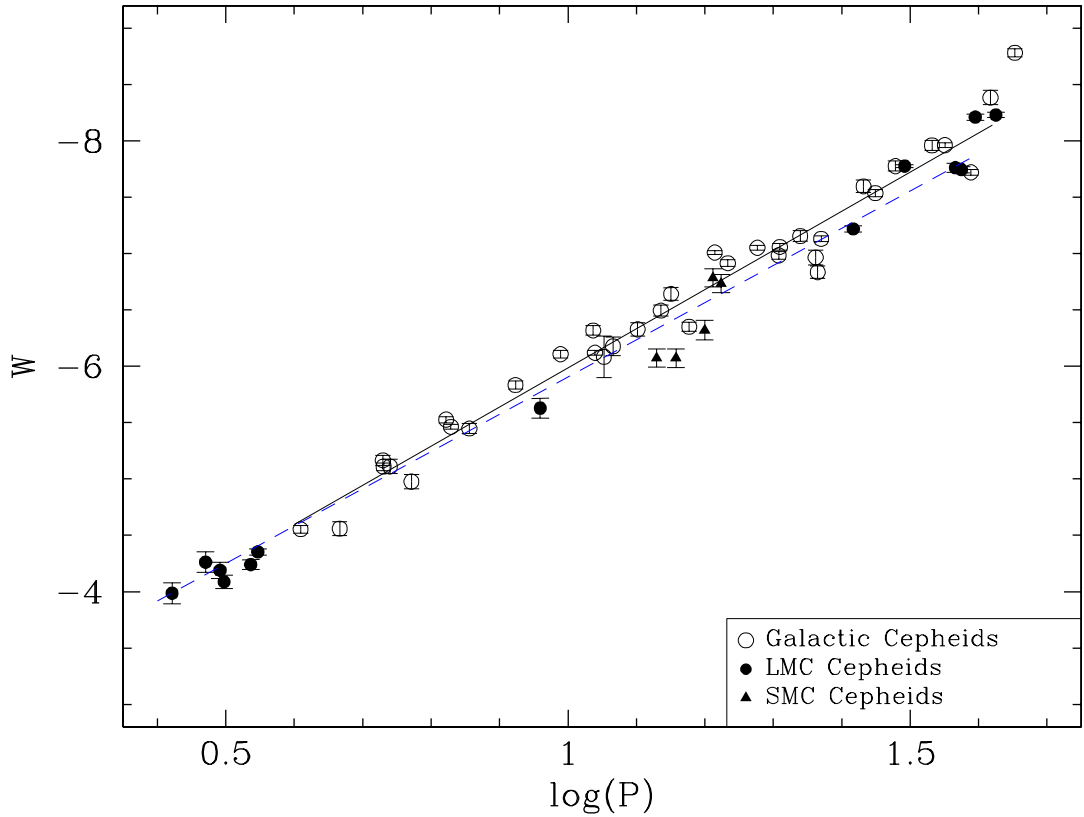}
\caption{The absolute Wesenheit magnitudes, calculated with the revised p-factor law. The solid
line is the best fit to the Milky Way Cepheid data. The dashed line is the OGLE-II relation
observed for the LMC, for an assumed LMC distance modulus of 18.50. The slopes of the W-band PL relations
for both the LMC and Milky Way from the ISB technique agree now very well with the observed
W-band relation in the LMC.}
\end{figure}

\end{document}